\documentclass[onecolumn]{aa} 
\usepackage[varg]{txfonts}
\usepackage{hyperref}
\usepackage{graphicx}
\usepackage{multirow}

\bibpunct{(}{)}{;}{a}{}{,}

\begin{document}

\title{Electron-beam interaction with emission-line clouds in blazars}

\author{Christoph Wendel\inst{1} \and Josefa Becerra Gonz\'{a}lez\inst{2} \and David Paneque\inst{3} \and Karl Mannheim\inst{4}}
\institute{Universit\"at W\"urzburg, D-97074 W\"urzburg, Germany, \email{cwendel@astro.uni-wuerzburg.de}
\and Inst. de Astrof\'isica de Canarias, E-38200 La Laguna, and Universidad de La Laguna, Dpto. Astrof\'isica, E-38206 La Laguna, Tenerife, Spain, \email{jbecerra@iac.es}
\and Max-Planck-Institut f\"ur Physik, D-80805 M\"unchen, Germany, \email{dpaneque@mppmu.mpg.de}
\and Universit\"at W\"urzburg, D-97074 W\"urzburg, Germany, \email{mannheim@astro.uni-wuerzburg.de}}

\abstract
{An electron-positron beam escaping from the magnetospheric vacuum gap of an accreting black hole interacts with recombination-line photons from surrounding gas clouds. Inverse-Compton scattering and subsequent pair production initiate unsaturated electromagnetic cascades exhibiting a characteristic spectral energy distribution.}
{By modelling the interactions of beam electrons (positrons) with hydrogen and helium recombination-line photons, we seek to describe the spectral signature of beam-driven cascades in the broad emission-line region of blazar jets.}
{Employing coupled kinetic equations for electrons (positrons) and photons including an escape term, we numerically obtain their steady-state distributions and the escaping photon spectrum.}
{We find that cascade emission resulting from beam interactions can produce a narrow spectral feature at TeV energies. Indications of such an intermittent feature, which defies an explanation in the standard shock-in-jet scenario, have been found at $\approx\,4\,\sigma$ confidence level at an energy of $\approx 3$\,TeV in the spectrum of the blazar Mrk~501.}
{The energetic requirements for explaining the intermittent 3\,TeV bump with the beam-interaction model are plausible: Gap discharges that lead to multi-TeV beam electrons (positrons) carrying $\approx0.1 \, \%$ of the Blandford-Znajek luminosity, which interact with recombination-line photons from gas clouds that reprocess $\approx 1\,\%$ of the similar accretion luminosity are required.}


\maketitle

\section{Introduction}\label{SectionIntroduction}

The launching of astrophysical jets from accreting black holes (BHs) involves fundamental gravitomagnetic processes. The frame-dragging effect on the plasma of the inner accretion flow and its advected magnetic field creates a rotating magnetosphere that can lead to a vacuum voltage drop of $10^{20} \, \rm{eV}$ for a typical spinning supermassive BH in the centres of galaxies \citep{LR}. It is commonly thought that thermal pair production (PP) in photon-photon collisions provides charge carriers in sufficient numbers to supply the Goldreich-Julian density, which allows electric currents to remove the space charges resulting from the electric potential and to reshape the magnetosphere into a force-free configuration. Any imbalance in the supply of charged particles, however, results in the immediate creation of vacuum (electrostatic) gaps. In vacuum gaps the actual charge density, provided by photon-photon PP \citep[see, however,][for an assessment of the importance of triplet PP in limiting the gap growth]{Petropoulou}, is different from the Goldreich-Julian charge density. Because of this deviation, the force-free condition is not satisfied and the electric field component parallel to the magnetic field lines is non-vanishing and able to accelerate intruding charge carriers to relativistic velocities, extracting energy from the central rotating BH \citep{BlandfordZnajek,Beskin,1998Hirotani,2000PhRvL..85..912L}.

The interaction of a particle beam with a low-energy radiation field results in electromagnetic cascades. Inverse-Compton (IC) scattering of the primary electrons off low-energy photons produces gamma rays that then undergo PP by interacting with the same photon field leading to a secondary generation of electrons. This sequence repeats until the gamma rays can escape freely \citep{Lovelace,1985Akharonian,Svensson,Zdz,Aharonian}.

Cascades have been studied in a variety of astrophysical environments. In outer magnetospheric vacuum gaps of pulsars, curvature radiation is thought to initiate the pair cascade, limiting the gap growth by supplying charged particles \citep[cf.][]{CHR2,Hirotani,Tang}. The gamma rays produced by the cascade and leaving the pulsar magnetosphere in a confined beam are made responsible for the observed high-energy (HE) and very high energy (VHE) pulsed emission. In the galactic centre, vacuum gaps are hypothesised to drive cascades that are responsible for PeV cosmic rays and for the VHE gamma-ray point source coincident with Sagittarius A* \citep[cf.\ e.g.][]{LR,2020ApJ...899L...7K}.

In active galactic nuclei (AGN) with high accretion rates such as Seyfert galaxies,  IC pair cascades in the coronal plasma of the accretion disk are thought to shape their X-ray spectrum up to MeV energies \citep{Svensson,Zdz}. In AGN with jets such as blazars, unsaturated IC pair cascades can develop from which gamma rays at energies well above MeV can escape. The cascades can be initiated by relativistic particles accelerated in vacuum gaps in the BH magnetosphere when the accretion rate is too low to supply the Goldreich-Julian density but their magnetic field is strong enough to launch a jet \citep[][for a review]{Neronov2,Neronov,LR,2017Hiro,2020ApJ...895...99K,2018Galax...6..122H}. The gaps are thought to be located at the poles of the spinning BH magnetosphere near to the null surface where the Goldreich-Julian charge density is equal to zero \citep{Beskin,1998Hirotani,Ptitsyna,HirotaniPu2,Levinson,Ford,Chen2, 2020arXiv200702838K}. It was also found that gaps form around the divide of the magnetohydrodynamic (MHD) flow, called the stagnation surface \citep{2010Vincent,Broderick,HirotaniPu1,AharonianVariability}, or near to the inner light surface of the magnetosphere \citep{Crinquand}. Time-dependent one-dimensional, general relativistic MHD simulations by \citet{Levinson}, \citet{LevinsonCerutti}, \citet{Chen}, and \citet{Chen2} and two-dimensional simulations by \citet[][for computational reasons, the latter authors had to restrict themselves to unrealistic parameters, however]{Crinquand} have recently found inherently non-stationary gap solutions. \citet{LevinsonCerutti} and \citet{2020arXiv200702838K} indicate that the gap activity relaxes to low-amplitude quasi-steady oscillations after an initial spark event, while \citet{Chen} and \citet{Chen2} find cyclically enduring gap activity.

Gap-accelerated particles then interact with ambient photon fields, that is, photons from the accretion flow or from the broad line region (BLR), initiating the onset of a cascade \citep[cf.][]{BlandfordLevinson,Broderick}, which gives rise to an additional component of IC up-scattered photons that are emitted along a beam in the same direction as the original electrons were moving. When these cascaded photons can escape from the AGN gamma sphere without being absorbed by PP, and when the line of sight coincides with the beam, the emission could cause an observable imprint on the spectral energy distribution (SED) of the AGN\@. Intermittency of the particle flux injected from the gap or of the target photon field results in time variability of the emerging radiation and in transient features of the total SED\@. This could serve as an explanation of the observed minute- and hour-scale variability of HE and VHE emission in blazars \citep[cf.][]{ReferenceVariability1,ReferenceVariability2,ReferenceVariability3,ReferenceVariability5,ReferenceVariability7}. 
Similar short-term variability has been found in AGN with jets at inclination angles larger than a few degrees, which supports a similar interpretation \citep[cf.][]{ReferenceVariability4,ReferenceVariability6}.
Alternative scenarios to explain the short-time variability are the interaction of the jet with the radiation field from a stellar envelope, for instance, or a relativistically propagating emission zone in a relativistically moving jet \citep[cf.][]{AharonianVariability,RiegerVariability}.

In this paper, we examine IC pair cascades in blazars emphasising the interaction of beam electrons from (short-lived) vacuum gaps with recombination-line photons. Blazars are subdivided into two subclasses, flat-spectrum radio quasars (FSRQs) and BL Lacertae (BL Lac) objects \citep[for a recent review, see e.g.][]{BlazarReview}. These two types were originally distinguished by the presence of broad  optical emission lines in the case of FSRQs, with equivalent width $|W_{\lambda}| > 0.5 \, \rm{nm}$, and the absence of broad optical emission lines for BL Lac objects \citep{Urry}. However, even in BL Lac objects, in which by definition $|W_{\lambda}| < 0.5 \, \rm{nm}$, optical and ultraviolet recombination lines from the BLR are often present \citep[cf.][for Markarian~501 and S4~0954+65, respectively]{Stocke,PepaInPrep}, and might imprint a diagnostically important signature on the gamma-ray spectra emerging from the innermost regions.

This paper is organised as follows: In Sect. \ref{SectionPairCascadeModel} we describe the physical model for the beam-photon interaction scenario and its underlying assumptions, including the set of coupled kinetic equations that were employed to numerically evaluate the model. We outline the numerical procedure for solving them to obtain the steady-state particle and photon distributions in Sect. \ref{SectionNumericalSolutionProcedure}. 
In Sect. \ref{SectionInverseComptonPairCascadeInMarkarian501} we apply this numerical scheme to fit the observational SED of the BL Lac object Markarian~501 (Mrk~501), which exhibited a peculiar spectral feature at $\approx 3$~TeV during a strong flaring activity in 2014 July \citep{JBG_Mrk501_bump}, and discuss the implications of the proposed scenario for the jet formation in Mrk~501. Finally, Sect. \ref{SectionSummary} summarises our findings.

In the following $m_{\rm e}$, $c$, $e$, $\sigma_{\rm Th}$, $k_{\rm B}$, and $h$ denote the electron rest mass, the velocity of light, the elementary charge, the Thomson cross section, the Boltzmann constant, and Planck's constant, respectively.  SI units are employed unless noted otherwise.

\section{Unsaturated inverse-Compton pair cascade model}
\label{SectionPairCascadeModel}
The electron beam from an active spark gap region close to the accreting BH can initiate IC pair cascades \citep{Broderick,Ptitsyna,Ford,LevinsonCerutti,Chen2}.
When the beam electrons propagate further along the jet axis, they can
encounter the intense photon field of an emission-line cloud, as depicted in Fig.~\ref{FigureGeometrySketch2}, and initiate IC pair cascades.
Because the threshold energy for PP in photon-photon collisions with the optical and ultraviolet photons characteristic for recombination-line spectra is high, the cascade spectra do not saturate at the electron rest mass energy delivering X-ray spectra, but instead produce spectra peaking at VHE gamma rays. Such cascades are also linear because the reservoir of low-energy target photons is not modified by the cascade in a major way.  For their SEDs, it is essential to consider an escape term accounting for the energy-dependent gamma sphere resulting from the pair creation optical depth, which decreases towards the edge of the emission region. 

\subsection{Unsaturated inverse-Compton pair cascade with escape}
\label{SubsectionModelAssumptions}

For the model of the emission region, we treat electrons and positrons identically and assume three reservoirs of particle species to be present:
\begin{itemize}
    \item {Relativistic electrons with spectral number density $N(\gamma)$, where $\gamma$ denotes the Lorentz factor of the electron.}
    \item {Highly energetic photons\footnote{In what follows, the term "highly energetic photon" does not only denote photons with energies in the established HE range from 100 MeV to 100 GeV, but more generally photons with energies well above $m_{\rm e} c^2$, in contrast to low-energy photons with energies well below $m_{\rm e} c^2$.} with spectral number density $n_\gamma(x_\gamma)$, where $x_\gamma$ denotes the photon energy in units of the electron rest mass energy $m_{\rm e}c^2$.}
    \item {External low-energy photons with spectral number density $n_0(x)$ with $x_1 \le x \le x_0 \ll 1$. }
\end{itemize}

\begin{figure*}
\parbox{.25\linewidth}{
\includegraphics[width=0.25\textwidth]{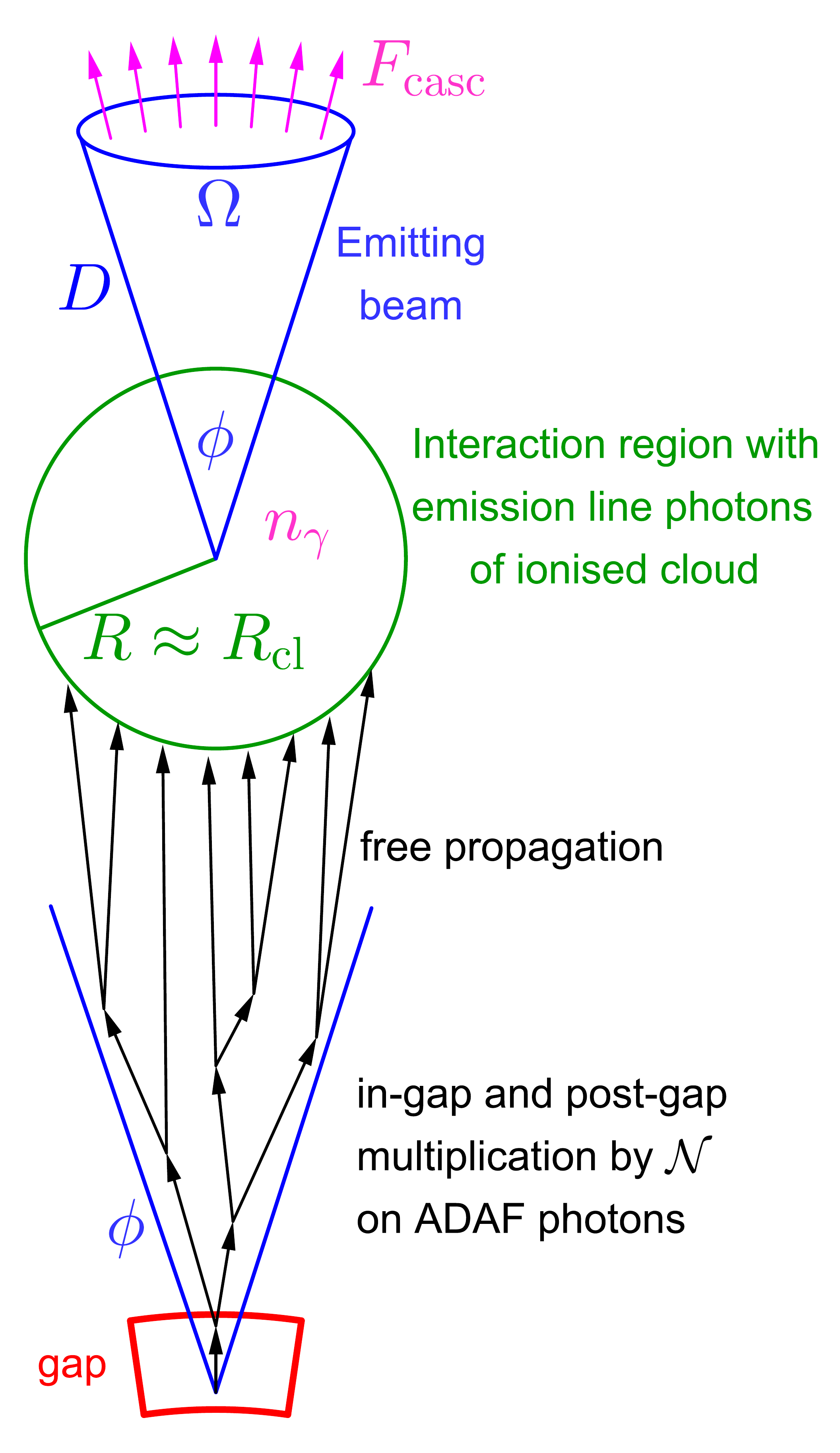}
}
\parbox{.19\linewidth}{
\vspace{-0.8cm}
\caption{Sketch of the geometry (not to scale) assumed for the model. Seed electrons are accelerated in the gap and multiplied in and behind the gap through ADAF photons. For simplicity of drawing, photons are not shown in the in-gap and post-gap multiplication. After propagating away from the gap, the electron beam meets an emission-line photon field, driving a cascade, whose steady-state gamma-ray photon density is $n_{\gamma}$ and whose escaping flux is $F_{\rm{casc}}$. To convert $n_{\gamma}$ into $F_{\rm{casc}}$, cf.\ Eq.~\ref{EquationCascadedFlux}.}
\label{FigureGeometrySketch2}
}
\hfill
\parbox{.48\linewidth}{
\includegraphics[width=0.48\textwidth]{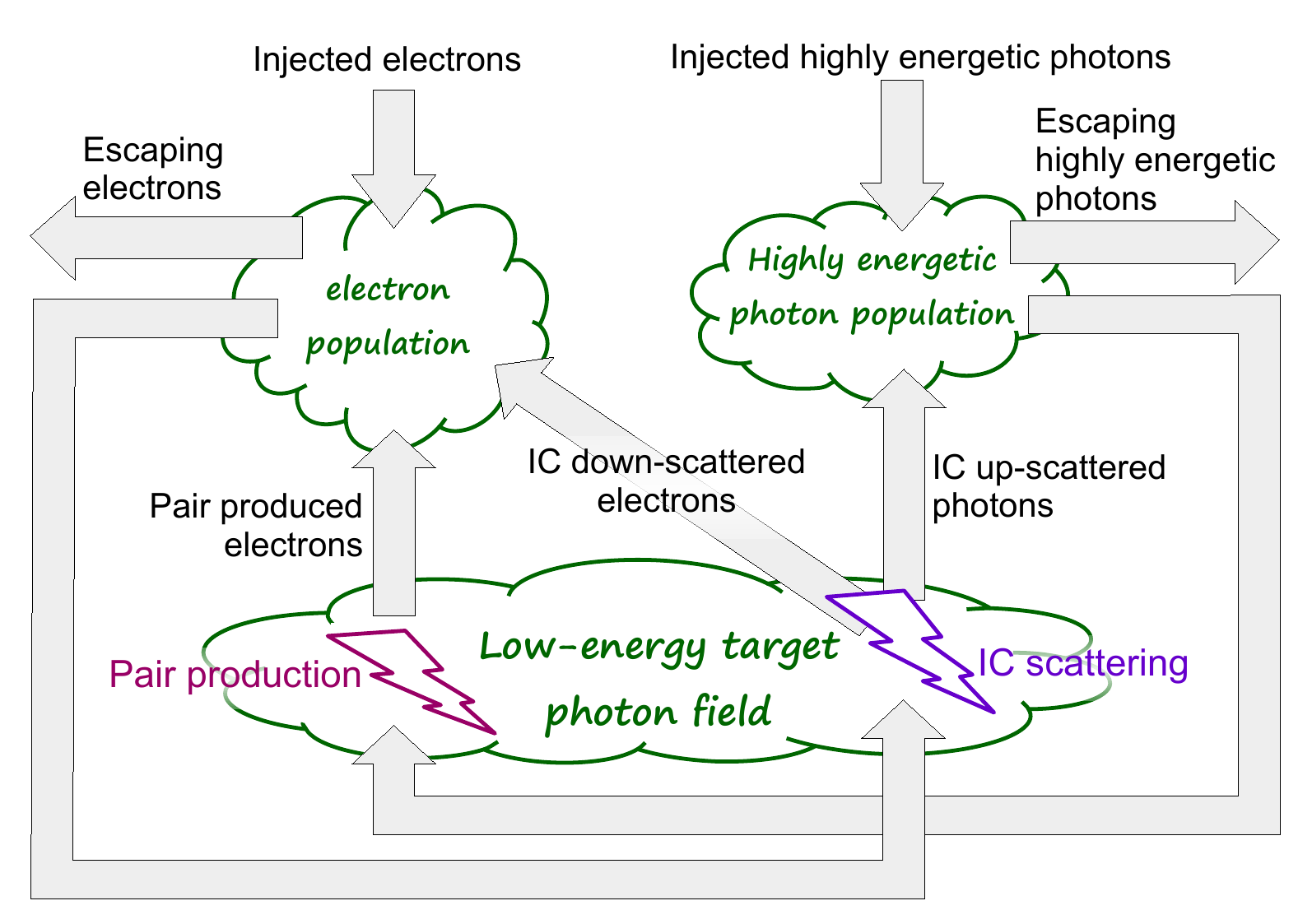}
\caption{Sketch of the considered linear IC pair cascade with escape terms.}
\label{FigureCascadeSketch}
}
\end{figure*}

We assume that all number densities are time-independent, isotropic, and homogeneous throughout the interaction region, and that interactions take place only inside of it. The interactions are assumed to be dominated by IC scattering and photon-photon PP. The cascade is then driven by PP from collisions in which the low-energy target photons turn the highly energetic photons into electron-positron pairs that then IC scatter off the same low-energy target photons to become the next generation of highly energetic photons, and so on (cf.\ Fig.~\ref{FigureCascadeSketch}).
We consider four additional mechanisms acting as sinks or sources for the highly energetic photons or the electrons:
\begin{itemize}
	\item {The injection of highly energetic photons with spectral rate $\dot n_{\gamma,\,\rm i}(x_\gamma)$ and $x_\gamma \gg 1$\footnote{$\dot n_{\gamma,\,\rm i}$ and $\dot N_{\rm i}$ are called spectral injection rate henceforth, and are defined as the number of particles, injected per unit space volume interval, per unit time interval, and per unit energy interval.\label{fnote2}}. Additionally, we demand that the threshold condition for PP in collisions with the isotropic target radiation field with $x_\gamma \, x > 1$ is fulfilled.}
	\item {The injection of electrons with spectral rate $\dot N_{\rm i}(\gamma)$ \textsuperscript{\ref{fnote2}}. We require that $\dot N_{\rm i}$ is non-vanishing only for $\gamma \gg 1$, hence the electrons have to be highly relativistic. Furthermore, we require that $\gamma \, x > 1$ is satisfied for the injected electrons. This is equivalent with primary IC scattering events being in the Klein-Nishina (KN) regime. According to \citet{Zdz}, this type of cascade is denoted IC-Klein-Nishina pair cascade.}
	\item {The escape of highly energetic photons from the interaction region with timescale $T_{\rm{ph\,esc}}(x_\gamma)$.}
	\item {The electron escape with timescale $T_{\rm{e\,esc}}(\gamma)$. This is the main difference of our treatment to that by \citet{Zdz} and \citet{Wendel17}, who considered saturated cascades, and makes our scenario more realistic. A finite escape timescale is equivalent to a bounded interaction region and is thus closer to reality than an infinite escape timescale, which is equivalent to an infinitely large interaction volume. The cascades considered here are therefore not necessarily saturated.}
\end{itemize}

We assume that the reservoir of target photons is not modified by the cascade. In other words, $n_0(x)$ is kept constant. This assumption is equivalent to assuming that a cascade is linear \citep{Zdz}. Linearity of a cascade means that the low-energy photons alone serve as targets. Self-interactions of the highly energetic photons and of the electrons are neglected.
\citet{Zdz} showed by comparing cross sections that when the energy density of low-energy photons is at least similar to that of the highly energetic photons, the condition for a cascade to be linear is met whose interactions occur in the KN regime. Repetition of IC scattering events successively decreases the electron energy, until IC scattering events take place in the Thomson regime ($\gamma \, x < 1$). Here, cascades are always non-linear because $x_\gamma \, x < 1$ implies that PP is not possible and thus PP is only possible through highly energetic photon self-interactions.

\subsection{Description through kinetic equations}\label{SubsectionDescriptionViaKineticEquations}

To determine the change rates of the spectral number densities of photons and electrons, we use the spectral interaction rates of IC scattering and of PP given by \citet{Zdz}.

For IC events with incident electrons of energy $\gamma \gg 1$ scattering off photons with energy $x \ll 1$ and spectral number density $n_0(x)$, which result in down-scattered electrons with energy $\gamma'$ and highly energetic photons with energy $x_\gamma$, the spectral IC scattering interaction rate is denoted by $C(\gamma,\gamma')$ and determined with Eq.~A1 by \citet{Zdz}, cf.\ Appendix \ref{AppendixSectionDescriptionViaKineticEquations}. For such events, the threshold $\gamma_{\rm{IC,\,th}}(x_\gamma,x)$ as defined in Appendix \ref{AppendixSectionDescriptionViaKineticEquations} is the minimum required value of $\gamma$. The maximum possible $\gamma$ is $\gamma_{\rm{IC,\,max}}(\gamma',x)$, as outlined in Appendix \ref{AppendixSectionDescriptionViaKineticEquations}. Furthermore, we define $\gamma'_{\rm{IC,\,min}}(\gamma,x)$ as the minimum possible $\gamma'$ and $x_{\gamma,\,\mathrm{max}}(\gamma,x)$ as the maximum energy of the up-scattered photon.

Given the interaction of highly energetic photons of energy $x_\gamma$ with low-energy photons of energy $x$ and spectral number density $n_0(x)$, resulting in the PP of electrons of energy $\gamma$, the spectral PP interaction rate is called $P(x_{\gamma},\gamma)$ and is computed with Eq.~B1 of \citet{Zdz}, cf.\ Appendix \ref{AppendixSectionDescriptionViaKineticEquations}. The PP threshold, that is, the minimum required $x_\gamma$, is defined as $x_{\gamma,\,\rm{PP,\,th}}(\gamma,x)$ in Appendix \ref{AppendixSectionDescriptionViaKineticEquations}. The values of $\gamma$ that can be reached are limited by $\gamma_{\rm{PP,\,min}}(x_\gamma,x)$ and $\gamma_{\rm{PP,\,max}}(x_\gamma,x)$, as described in Appendix \ref{AppendixSectionDescriptionViaKineticEquations}.

The kinetic equation of the electrons and of the photons ensues after all the relevant sinks and sources are specified. The change rate of the spectral number density of the electrons (electron energy distribution) is affected by the electron spectral injection rate $\dot N_{\rm i}$, the spectral loss rate (as the sum of the escape rate with timescale $T_{\rm{e\,esc}}$ and of the IC down-scattering rate to lower energies), and by the production rate (as the sum of the IC down-scattering rate from higher energies and of the PP rate). Similarly, the change rate of the spectral number density of the photons is determined from the photon spectral injection rate $\dot n_{\gamma,\,\rm i}$, the loss rate (which is the sum of the escape rate with timescale $T_{\rm{ph\,esc}}$ and of the pair absorption rate), and the production rate $\dot n_{\gamma, \, \rm{IC}}$.

Assuming steady state, we obtain, as shown in Appendix \ref{AppendixSectionDescriptionViaKineticEquations},
\begin{equation}
n_\gamma(x_\gamma) = 
\underbrace{\frac{\dot n_{\gamma,\,\mathrm{i}}(x_\gamma)}{\frac{1}{T_{\rm{ph\,esc}}(x_\gamma)} + \int ^{\gamma_{\mathrm{PP,\,max}}(x_\gamma,x_0)}_{\gamma_{\mathrm{PP,\,min}}(x_\gamma,x_0)} P(x_\gamma,\gamma) \, \mathrm{d}\gamma}}_{:= \;n_{\gamma,\,\mathrm{i}}(x_\gamma)} +
\underbrace{\frac{\int _{\gamma_{\mathrm{IC,\,th}}(x_\gamma,x_0)}^{\infty} N(\gamma) C(\gamma,\gamma-x_\gamma) \, \mathrm{d}\gamma}{\frac{1}{T_{\rm{ph\,esc}}(x_\gamma)} + \int ^{\gamma_{\mathrm{PP,\,max}}(x_\gamma,x_0)}_{\gamma_{\mathrm{PP,\,min}}(x_\gamma,x_0)} P(x_\gamma,\gamma) \, \mathrm{d}\gamma}}_{:= \; n_{\gamma,\,\mathrm{IC}}(x_\gamma)},
\label{EquationSpectralNumberDensityHEPs}
\end{equation}
where we have defined two contributions to the photon spectral number density. $n_{\gamma,\,\mathrm{i}}$ describes the contribution that arises from the interplay of injection and escape and absorption losses, while $n_{\gamma,\,\mathrm{IC}}$ describes the contribution that is achieved from the interplay of production through IC up-scattering and both losses. Exchanging $\gamma$ and $\gamma'$ in Eq.~\ref{EquationSpectralNumberDensityHEPs} and plugging this into Eq.~\ref{EquationSpectralNumberDensityElectrons}, we can eliminate $n_\gamma$,
\begin{align}
&N(\gamma) = \label{EquationKineticEquation}\\
&\frac{\dot N_{\mathrm{i}}(\gamma) + \int _{\gamma}^{\gamma'_{\mathrm{IC,\,max}}(\gamma,x_0)} N(\gamma') C(\gamma',\gamma) \, \mathrm{d}\gamma' + \int _{x_{\gamma,\,\mathrm{PP,\,th}}(\gamma,x_0)}^{\infty} \left( \dot n_{\gamma,\,\mathrm{i}}(x_\gamma) + \int _{\gamma'_{\mathrm{IC,\,th}}(x_\gamma,x_0)}^{\infty} N(\gamma') C(\gamma',\gamma'-x_\gamma) \, \mathrm{d}\gamma' \right) \frac{2 \: P(x_\gamma,\gamma)}{\frac{1}{T_{\rm{ph\,esc}}(x_\gamma)} + \int ^{\gamma'_{\mathrm{PP,\,max}}(x_\gamma,x_0)}_{\gamma'_{\mathrm{PP,\,min}}(x_\gamma,x_0)} P(x_\gamma,\gamma') \, \mathrm{d}\gamma'} \, \mathrm{d}x_\gamma}{\frac{1}{T_{\rm{e\,esc}}(\gamma)} + \int ^{\gamma}_{\gamma'_{\rm{IC,\,min}}(\gamma,x_0)} C(\gamma,\gamma') \, \mathrm{d}\gamma'} \nonumber
\end{align}

\section{Numerical solution procedure}\label{SectionNumericalSolutionProcedure}

In this section, we outline how explicit solutions for $N(\gamma)$ and $n_\gamma(x_\gamma)$ are found.

The main task is to find solutions for $N$. This is done by solving Eq.~\ref{EquationKineticEquation} iteratively. The right-hand side of Eq.~\ref{EquationKineticEquation} directly incorporates $\gamma$, $T_{\rm{ph\,esc}}$, $T_{\rm{e\,esc}}$, $N$, $\dot N_{\rm{i}}$, $\dot n_{\gamma,\,\rm i}$, $C$, $P$, and the functions in the integration borders, which have been defined in the previous section. $C$ and $P$ are known as soon as $n_0$ has been specified. Hence, Eq.~\ref{EquationKineticEquation} can briefly be written as $N(\gamma) = \mathcal{F}'(n_0, \dot N_{\rm i}, \dot n_{\gamma,\,\rm i}, T_{\rm{ph\,esc}}, T_{\rm{e\,esc}}, N, \gamma)$. As described in Sect. \ref{SubsectionModelAssumptions}, the first five arguments of $\mathcal{F}'$ are quantities that describe the physical setting. Hence, they should be understood as input functions and have to be prescribed to specify the physical problem. When this is done, we obtain the equation $N(\gamma) = \mathcal{F}(N, \gamma)$, where $\mathcal{F}$ was defined accordingly. Now, we choose a sequence $\left( \gamma_k \right)_{k = 1,\dots,\kappa}$ of $\kappa$ discrete values $\gamma_k$ along which the kinetic equation is to be solved, in other words, along which $N(\gamma)$ is to be determined. We write $N(\gamma_k) = N_k$ for convenience. As a starting point of the iteration, we guess an initial course $N_{\rm{init}}(\gamma)$ of the function $N(\gamma)$ and determine the corresponding sequence of values $\left( N_k \right)_{k = 1,\dots,\kappa;\,j_{\rm{init}}}$. In Appendix \ref{AppendixSectionNumericalSolutionProcedure}, we discuss how the converged sequence is determined numerically.

We perform the iteration using the python3 language. To numerically compute the integrals with an infinite integration range, we restrict the integration range we used to a finite range. This is possible when $n_0(x)$, $\dot n_{\gamma,\,\rm i}(x_\gamma)$, and $\dot N_{\rm i}(\gamma)$ are equal to zero everywhere above an upper cut-off, which we denote by $x_0$, $x_{\gamma,\,0}$, and by $\gamma_{{\mathrm{i}},\,0}$, respectively. 

To summarise, the physical setting is specified as soon as the following functions are defined:
\begin{itemize}
\item The low-energy photon spectral number density $n_0$
\item The highly energetic photon spectral injection rate $\dot n_{\gamma,\,\rm i}$
\item The electron spectral injection rate $\dot N_{\rm i}$
\item The highly energetic photon escape time $T_{\rm{ph\,esc}}$
\item The electron escape time $T_{\rm{e\,esc}}$
\end{itemize}
By iterating, our code then determines the steady state $N$. When $N$ is known, we use Eq.~\ref{EquationSpectralNumberDensityHEPs} to determine $n_\gamma$, and from this, we obtain the spectral flux density $F_{\rm{casc}}$. This approach differs from that of \citet{Wendel17}. Therein, although $T_{\rm{ph\,esc}}=T_{\rm{e\,esc}}=\infty$ (no escape) was used in the kinetic equation, it was assumed that the IC up-scattered photons leave the interaction region from its shell-like boundary. Accordingly, \citet{Wendel17} directly converted the photon spectral production rate (last term in Eq.~\ref{EquationRateOfChangeHEPs}) into the observed spectral flux density. This appears to be a very crude approach because first it neglects that the injected highly energetic photons can also leave the interaction region (which we take into account by $n_{\gamma,\,\mathrm{i}}$ in Eq.~\ref{EquationSpectralNumberDensityHEPs}), and second it assumes that all IC up-scattered photons escape from the boundary of the interaction region and that no IC up-scattered photons escape from its interior. On the one hand, this second assumption contradicts the $T_{\rm{ph\,esc}}=T_{\rm{e\,esc}}=\infty$ assumption and on the other hand it neglects the energy-dependent behaviour of photon attenuation (which we take into account through the division of the photon spectral injection and production rate (numerator of Eq.~\ref{EquationSpectralNumberDensityHEPs}) by the spectral loss rate (denominator of Eq.~\ref{EquationSpectralNumberDensityHEPs})).

Based on $n_\gamma$, we determine the spectral flux density $F_{\rm{casc}}$ of highly energetic photons (i.e.\ the number of photons per unit energy, per unit area, and per unit time) through
\begin{equation}
F_{\mathrm{casc}}(x_\gamma) = n_{\gamma}(x_\gamma) \cdot \frac{4 \pi \, c \,  R_{\rm{ph\,esc}}(x_\gamma)^2}{\Omega(\phi) \, D^2 \ m_{\rm{e}} c^2}.
\label{EquationCascadedFlux}
\end{equation}
We assumed that the photons leave the interaction region, whose approximate radial size is $R_{\rm{ph\,esc}}(x_\gamma)$, only through a conical beam of opening angle $\phi$ and solid angle $\Omega = 4 \pi \cdot \sin^2(\phi/4)$, and that $D$ is the distance to the observer. The size of the interaction region is linked with the escape time of the photons and electrons. A natural choice for this would be $R_{\rm{ph\,esc}}(x_\gamma) = T_{\rm{ph\,esc}}(x_\gamma) \, c$. The division by $m_{\rm{e}} c^2$ arises because $F_{\rm{casc}}$ is measured in $\rm{m}^{-2} \, \rm{s}^{-1} \, \rm{J}^{-1}$, while $n_{\gamma}$ is in units of $\rm{m}^{-3}$.

\section{Indication of gap activity in Markarian~501}\label{SectionInverseComptonPairCascadeInMarkarian501}

The numerical procedure described above is used below to explain a peculiar feature in the VHE spectrum of the blazar Mrk~501 during a flaring activity on 2014 July 19. 

The broadband emission from blazars, and in particular from the high-synchrotron-peaked BL Lac object Mrk~501, is characterised by a two-bump structure. This emission is usually described within the one-zone synchrotron-self-Compton (SSC) framework \citep[see e.g.][]{Maraschi,Dermer}.
In this model, a blob of relativistic electrons is assumed to move along the jet that points towards the observer. The blob is permeated by a randomly distributed magnetic field, producing synchrotron radiation from the electrons. This radiation is responsible for the low-energy SED bump in blazars. The origin of the HE bump is typically ascribed to the IC scattering of the relativistic electrons with the synchrotron photons. In addition to the SSC scenario, other theoretical models have been proposed to explain the HE and VHE emission in blazars, involving hadronic components and IC pair cascading. Minute-scale variability spectral components mean that the standard one-zone SSC scenario may be insufficient and suggest that an additional sporadic emission component is emitted from the innermost regions of the jet, see \citet{Ahnen18}.

Recently, the \citet{JBG_Mrk501_bump} has presented a two-week flare from Mrk~501, mainly detected in X-rays with the Swift satellite and in VHE gamma rays observed with the major atmospheric gamma-ray imaging Cherenkov (MAGIC) telescopes. During this flaring state, the source displayed hard X-ray spectra as well as the highest X-ray count rate detected by the Swift X-Ray Telescope during its 15 years of operation. In coincidence with the peak of the X-ray count rate, which took place on 2014 July 19 (MJD 56857.98), a narrow feature in the TeV spectrum from the MAGIC telescopes was detected with a significance of $3 \, \sigma - 4 \, \sigma$. This feature, located around $3 \, \rm{TeV}$, is inconsistent with the standard functions to describe the VHE spectra from blazars (power law, log-parabola, and log-parabola with exponential cut-off) at more than $3\,\sigma$ confidence level. The VHE spectrum on 2014 July 19 is better explained assuming a narrow component (a narrow log-parabola or a Gaussian function) in addition to a broad log-parabola. Such a double function fit is preferred with respect to a single log-parabola at $3.2 \, \sigma - 4.5\, \sigma$, depending on the a priori assumptions \citep[details can be found in][]{JBG_Mrk501_bump}. This was the first time that a narrow feature (not consistent with the usual smooth spectral functions) was found with a significance higher than $3\,\sigma$ in the VHE spectrum of Mrk~501, and any TeV blazar in general. Indications for similar (but statistically insignificant) spectral features at $\approx 2\,\rm{TeV}$ can be found by inspection of flare spectra reported by the very energetic radiation imaging telescope array system (VERITAS) \citep[cf.\ Fig.~8 of][]{2011ApJ...727..129A} and the MAGIC collaboration \citep[cf.\ Fig.~10 of][]{2017A&A...603A..31A}.

Under the assumption that the spectral feature described above is real, three different theoretical scenarios were proposed by the \citet{JBG_Mrk501_bump} to reproduce the narrow spectral feature: a) a pileup in the electron energy distribution produced by stochastic acceleration, b) a two-zone SSC model, and c) a narrow emission from an IC pair cascade scenario induced by electrons accelerated in a magnetospheric vacuum gap. Details for the model describing scenario c) are provided in this work. In the following, we evaluate the model we fitted to the observational data to infer its implications and consequences.

\subsection{Specification of the modelled setting}\label{SubsectionSpecificationOfTheModelledSetting}

We assume a vacuum gap near to the poles of the Mrk~501 BH magnetosphere. The potential drop in the gap accelerates seed electrons to ultra-relativistic energies. The seed electrons enter the gap both through PP of gamma rays from the hot advection-dominated accretion flow (ADAF), and possibly through direct leakage from it \citep{Neronov}. Seed electrons that have been accelerated in the gap, leave it in the direction of the electric field (i.e.\ in the direction of the magnetic field), and propagate along a beam that is well collimated to an opening angle $\approx \phi$. During the propagation away from the gap, the seed electrons still encounter the ADAF photon field, and a post-gap cascade \citep{Broderick,Chen2} occurs and increases the number of electrons by repeated curvature-radiation and IC emission and PP events. This post-gap cascade ceases when curvature-radiation emission becomes negligible and when IC scattering and PP can no longer be sustained by ADAF photons. Several tens or hundreds of Schwarzschild radii away from the gap, the beam of electrons is assumed to enter a region in which background photons from ionised gas clouds are present. This induces a cascade in this interaction region. This model is depicted in Fig.~\ref{FigureGeometrySketch2}. For this cascade, the model of Sect. \ref{SectionPairCascadeModel} is applied. The electron beam from the gap serves as electron injection mechanism in our model. We describe the electron spectral injection rate with the following cut-off Gaussian:
\begin{equation}
\dot N_{\rm i}(\gamma) = \left\{
\begin{array}{ll}
\frac{K_{\rm{G}}}{\varsigma \sqrt(2 \pi)} \cdot \exp \left( -\frac{(\gamma-\gamma_{\mathrm{mean}})^2}{2 \, \varsigma^2} \right) & \mathrm{if} \; \gamma_{{\mathrm{i}},\,1} \leq \gamma \leq \gamma_{{\mathrm{i}},\,0} \mathrm{,} \\
0 & \mathrm{otherwise}
\end{array}
\right.
\label{EquationMrk501DistributionElectrons}
\end{equation}
Here, $\gamma_{\rm{mean}}$ denotes the mean Lorentz factor of the injected electrons, and $\varsigma$ parametrises the width of the distribution.
For the cut-off values we choose $\gamma_{{\rm{i}},\,1} = \gamma_{\rm{mean}} - 3.0 \, \varsigma$ and $\gamma_{{\rm{i}},\,0} = \gamma_{\rm{mean}} + 3.0 \, \varsigma$, satisfying the condition $\gamma \cdot x > 1$ (cf.\ Sect. \ref{SubsectionModelAssumptions}). The normalisation of the Gaussian, which describes the total number of electrons that are injected per unit time and per unit space volume, is called $K_{\rm{G}}$.

Furthermore, we assume that the external low-energy photons of the cascade are represented by emission-line photons from photo-ionised gas clouds in the surroundings of the Mrk~501 BH magnetosphere. Although Mrk~501, as a BL Lac object, does not have a prominent BLR, its BLR has clearly been confirmed observationally, possibly due to outshining by the boosted jet emission, due to a weak accretion rate, or due to the lack of ambient gas \citep[cf.][]{Moles,Stocke}. After they are triggered by minor galaxy mergers, interstellar gas clouds, consisting mainly of hydrogen and helium \citep{Wilms}, can migrate from the host galaxy into its centre and thus into the central part of Mrk~501. Radiation from the accretion flow or from OB-type stars in the dense galactic centre will inevitably ionise passing clouds. As an effect of recombination, emission-line photons will be abundant in the proximity of the clouds. The external low-energetic photon field can also stem from the envelope of a post-main-sequence star with low metallicity \citep[cf.\ e.g.][]{1997MNRAS.287L...9B,2010ApJ...724.1517B}. In any case, we assume that the interaction region is both penetrated by the electron beam and pervaded by emission-line photons, and consequently, that an IC pair cascade will develop there. It is natural to describe the spectral number density by a sum of Dirac delta distributions,
\begin{equation}
n_{0}(x) = K_{\rm{lines}} \cdot \sum _{i=1}^4 \frac{K_{{\rm{line}},\,i}}{x_{0,\,i}} \cdot \delta_{\mathrm{Dirac}} \left( x - x_{0,\,i} \right)
\label{EquationMrk501DistributionLEPs}
\end{equation}
Here, $x_{0,\,i} = h / (\lambda_{0,\,i} \, m_{\rm{e}} \, c)$, and $\lambda_{0,\,i}$ are the dimensionless energy, and the wavelength of the $i$-th line, respectively. $K_{{\rm{line}},\,i}$ gives the relative flux density contribution of the $i$-th line\footnote{$K_{{\rm{line}},\,i}$ gives the flux density of the emission line $i$ with respect to the flux density of a hypothetical hydrogen Balmer-$\beta$ line.}. Accordingly, $K_{{\rm{line}},\,i}/x_{0,\,i}$ describes the relative number density contribution of the $i$-th line. The parameter $K_{\rm{lines}}$ is a measure for the photon total number density. We restrict the sum to the three typically most important hydrogen lines plus the most important helium line. We compiled these lines and their respective $K_{{\rm{line}},\,i}$ based on the ultraviolet spectroscopic blazar observations by \citet{Pian} and on the synthetic photo-ionisation spectra by \citet{Abolmasov}. We list them in Table \ref{TableMrk501Lines}.

From an energetic point of view, the emission-line radiation is just a reprocessed fraction of the more luminous ADAF radiation. Nevertheless, we do not include the ADAF radiation in the target photon field. This is justified by the following geometrical argument. The emission of line photons in the ionised cloud is approximately isotropic. The interaction angles of injected electrons and photons are therefore equally distributed; the interaction angle is on average $90 \degr$. This also justifies the usage of the interaction rates $C$ and $P$. In contrast, for interactions of electrons with ADAF photons the collision angle would be at best $90 \degr$, and on average smaller than $90 \degr$. Thus, glancing collisions occur, and for such interactions, the interaction rates are lower than for right-angled collisions. This results in increased interaction rates for the line photons in comparison to ADAF photons, and thus we can neglect the latter. Moreover, except for the reprocessed photons from the ionised cloud, reprocessed ADAF photons might be scarce in the surroundings of the beam due to low ambient gas density in the BL Lac object, which means that they drop out as target photons as well. We have to mention furthermore, that ADAF spectra usually extend to higher energies and $x \ll 1$, which was assumed in the derivation of $C$ and $P$, is satisfied only very critically.

We do not have injection of highly energetic photons in this setting. Therefore we set the photon spectral injection rate
\begin{equation}
\dot n_{\gamma,\,\rm i}(x_\gamma) = 0.
\label{EquationMrk501DistributionHEPs}
\end{equation}

Concerning the choice of the escape timescales, we choose
\begin{equation}
T_{\rm{ph\,esc}}(x_\gamma) = T_{\rm{e\,esc}}(\gamma) = \frac{R}{c} := T_{\rm{esc}}.
\label{EquationMrk501EscapeTimeScales}
\end{equation}
In other words, we approximate the interaction volume as a spherical region with radius $R$. Electrons and photons have equal, and energy independent, that is, constant, escape timescales. The function $R_{\rm{ph\,esc}}(x_\gamma)$ in Eq.~\ref{EquationCascadedFlux} is therefore also set equal to the parameter $R$.

\subsection{Modelling results}\label{SubsectionModellingResults}

According to the procedure outlined in Sect. \ref{SectionNumericalSolutionProcedure}, we determine $N(\gamma)$, $n_\gamma(x_\gamma)$, and $F_{\rm{casc}}(x_\gamma)$ of the cascade that evolves through interaction of the electrons originally stemming from the gap with the emission-line photons, such that $F_{\rm{casc}}(x_\gamma)$ is fitted to the narrow SED feature (MAGIC telescopes' data) and a background SSC emission is fitted to the multi-wavelength SED\@. We understand $K_{\rm{G}}$, $K_{\rm{lines}}$, $\gamma_{\mathrm{mean}}$, $\varsigma$, $R$, and $\phi$ as fitting parameters, while we use the luminosity distance $D = 149.4 \, \rm{Mpc}$ (corresponding to redshift $z=0.034$ and a cosmology\footnote{Here, $\Omega_{\rm{m}}$, $\Omega_{\Lambda}$, and $H_0$ are the dimensionless density parameter of matter, the dimensionless density parameter of dark energy, and the Hubble constant, respectively.} $\Omega_{\rm{m}} = 0.3$, $\Omega_{\Lambda} = 0.7$, $H_0 = 70 \, \rm{km} \, \rm{s}^{-1} \, \rm{Mpc}^{-1}$) and the line contributions $K_{{\rm{line}},\,i}$ given in Table \ref{TableMrk501Lines}. Because the electron beam penetrates into the interaction region from one direction, the kinematics of IC scattering and PP ensure that the highly energetic photons leave the interaction region in this same direction, hence along a beam of opening angle $\phi$.

\begin{figure*}
\parbox{.4\linewidth}{
\includegraphics[width=0.40\textwidth]{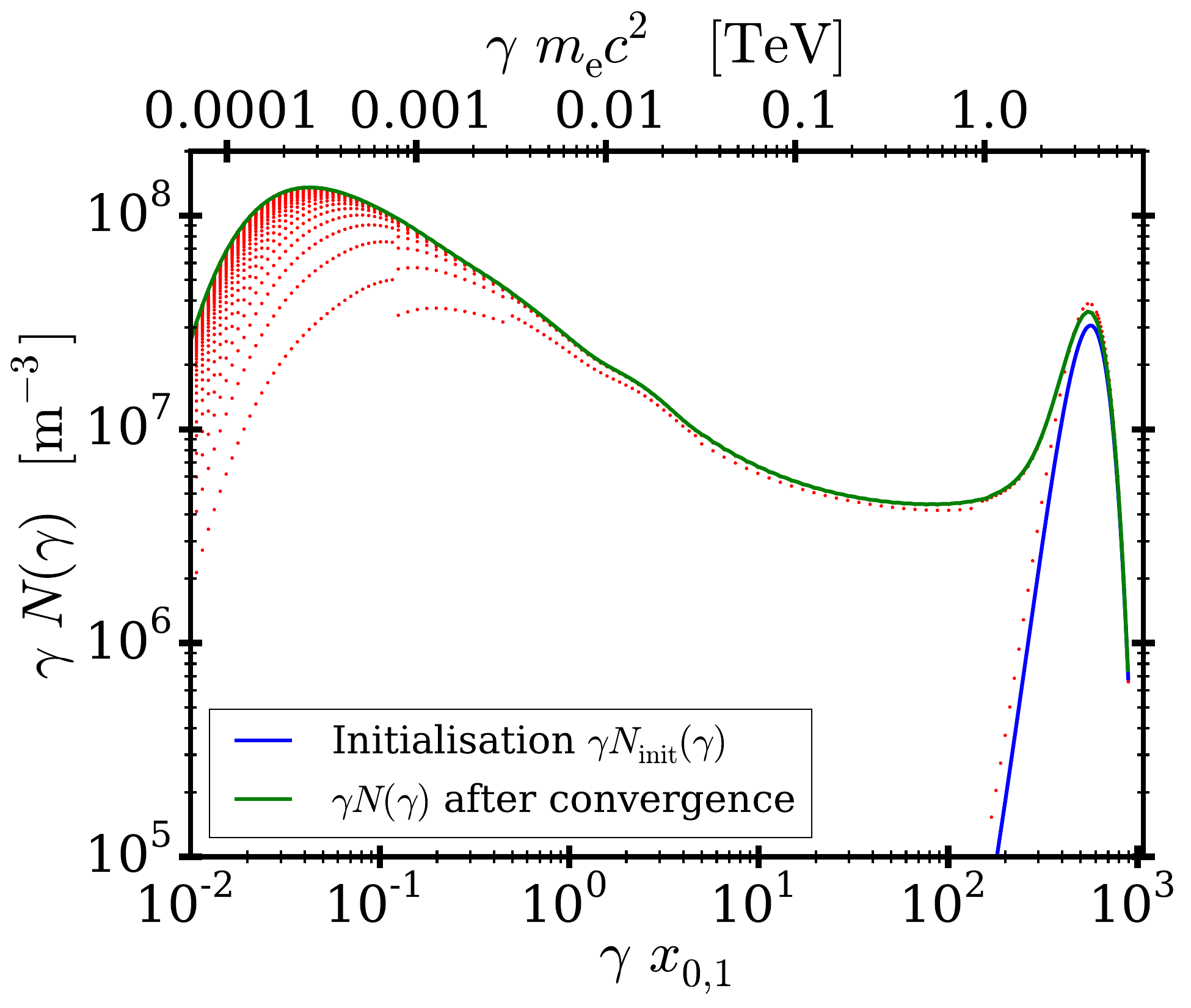}
\caption{$\gamma \, N(\gamma)$, which is a proxy for the electron spectral energy density, versus the product of the Lorentz factor with the dimensionless energy of the highest energetic line, the helium II Lyman-$\alpha$ line. The iterated points $N_{k,\,j}$ are drawn in red.}
\label{FigureNElectrons}
}
\hfill
\parbox{.57\linewidth}{
\includegraphics[width=0.57\textwidth]{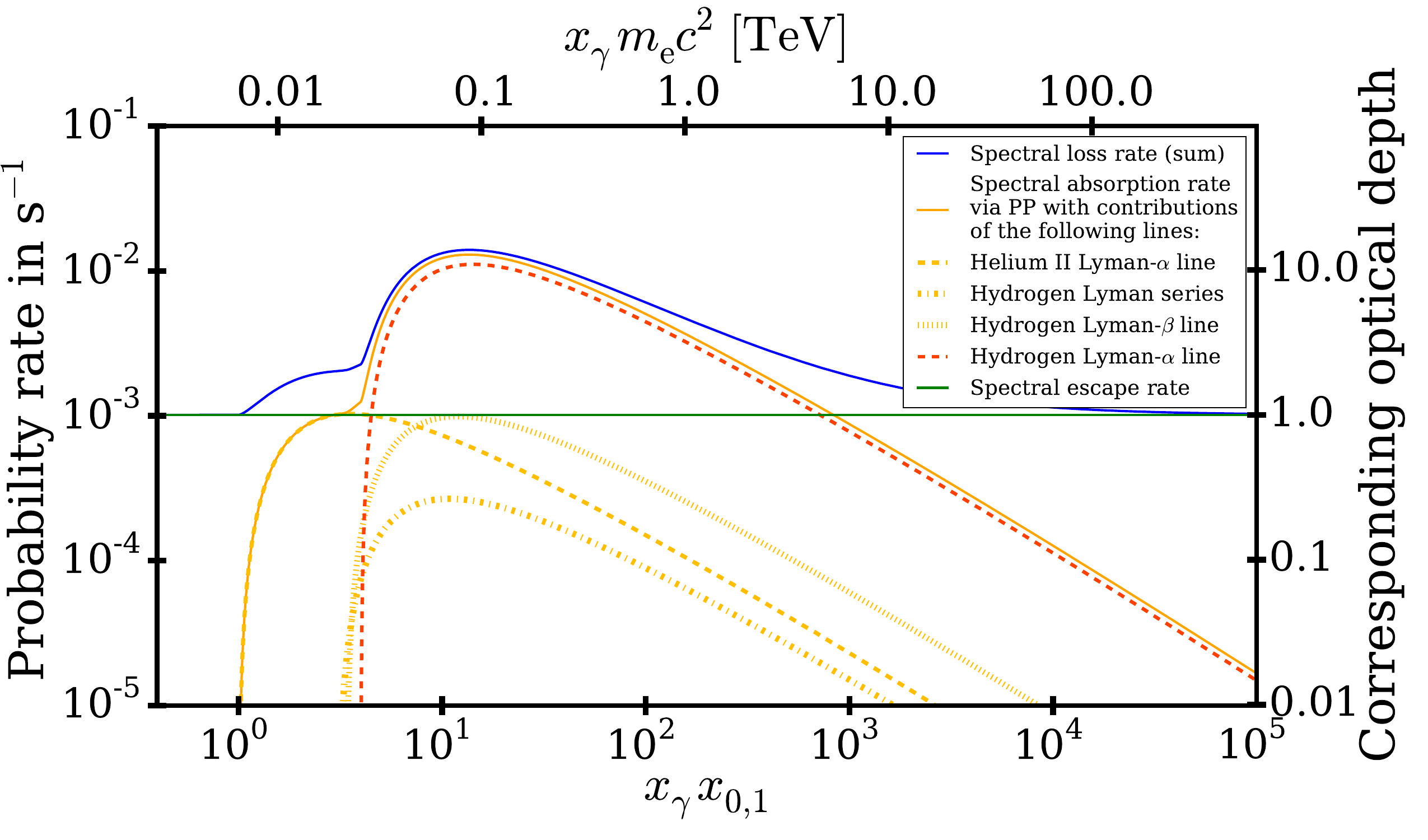}
\caption{Various contributions (cf.\ Eq.~\ref{EquationRateOfChangeHEPs}) to the photon spectral loss rate on the left-hand side ordinate in dependence on the product of the dimensionless highly energetic photon energy with the dimensionless energy of the highest energetic line. On the right-hand side ordinate, we plot the optical depth that corresponds to the respective contribution to the spectral loss rate.}
\label{FigurePPAndEscape}
}
\end{figure*}

\begin{figure*}
\parbox{.44\linewidth}{
\includegraphics[width=0.44\textwidth]{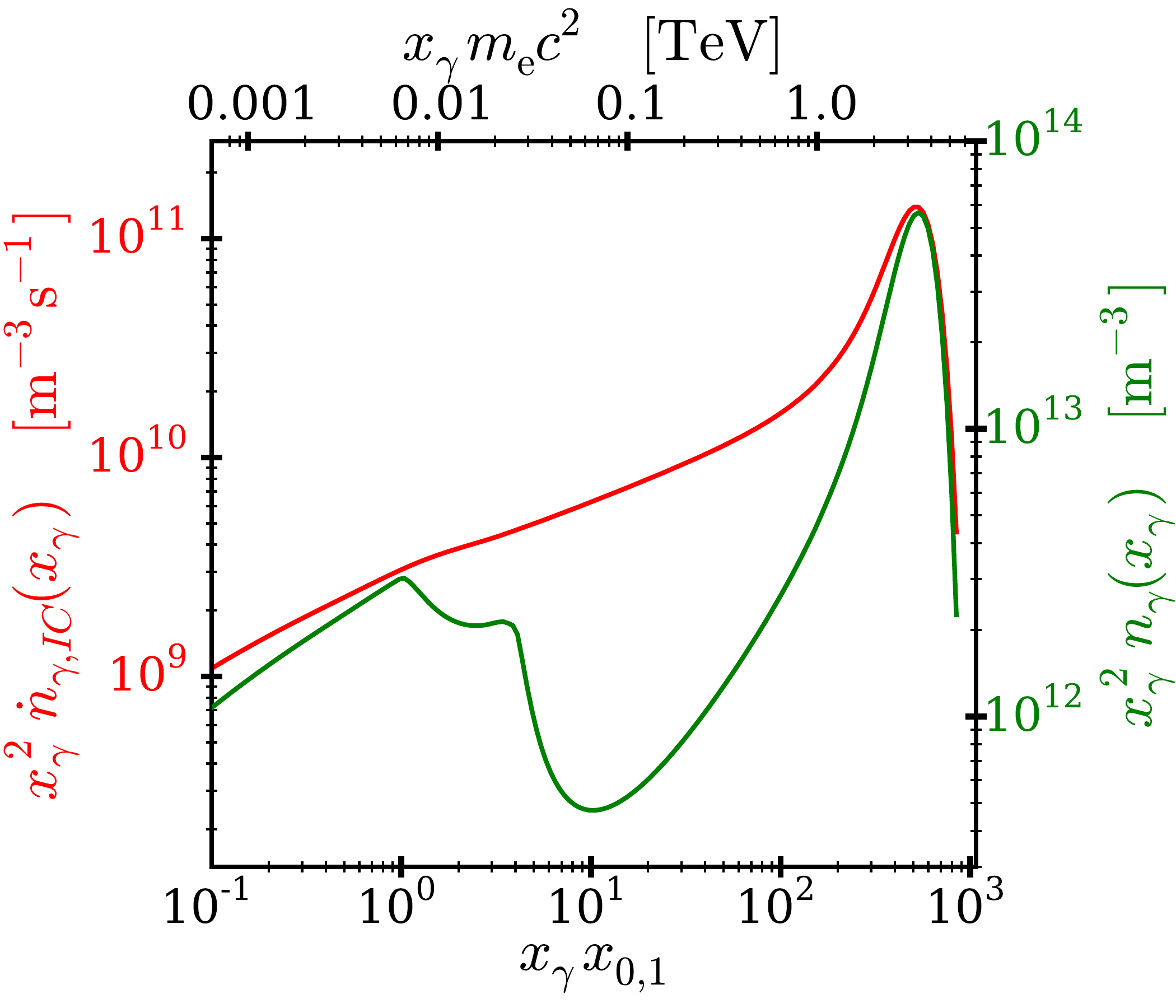}
\caption{Spectral production rate and number density of highly energetic photons of the cascade in Mrk~501. The product of the squared dimensionless photon energy with the spectral production rate of photons is plotted in red (which is determined by the last term of Eq.~\ref{EquationRateOfChangeHEPs}). We show the product of the squared dimensionless photon energy with the spectral number density of photons in green (which is determined by Eq.~\ref{EquationSpectralNumberDensityHEPs}). Both quantities are plotted versus the product of the highly energetic photon energy with the energy of the helium II Lyman-$\alpha$ line.}
\label{FigureDotnAndnPhotons}
}
\hfill
\parbox{.54\linewidth}{
\includegraphics[width=0.54\textwidth]{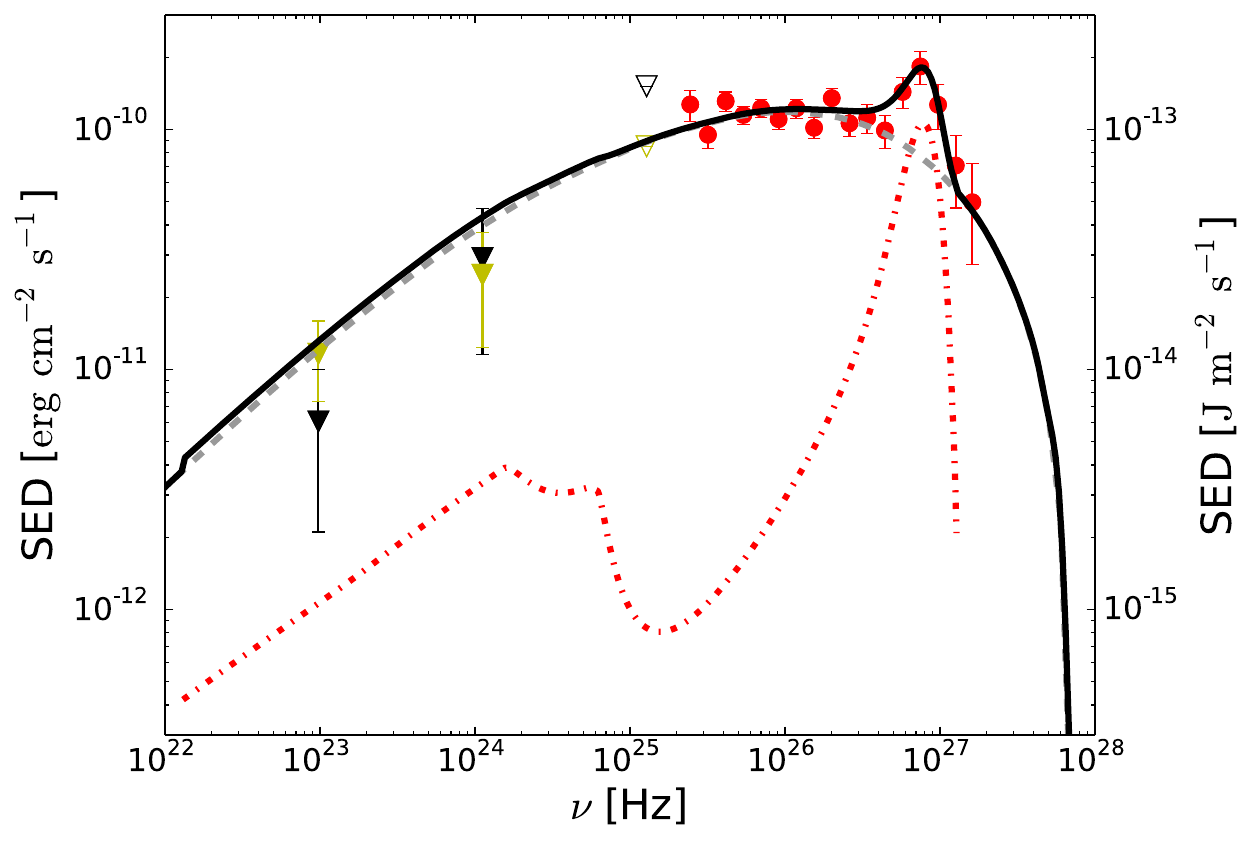}
\caption{High-energy bump of the SED of Mrk~501 from 2014 July 19 (MJD 56857.98). The dashed grey line depicts the IC bump of a one-zone SSC model (contribution $F_{\rm{SSC}}$). The dot-dashed red line depicts the emission of the cascade (contribution $F_{\rm{casc}}$, cf.\ green line in Fig.~\ref{FigureDotnAndnPhotons}). The sum $F$ of both contributions is depicted by the solid black line. The spectral data from the MAGIC telescopes are shown as red circles, while measurements and upper limits by the Fermi Large Area Telescope are drawn by black and yellow triangles. The details on the data analysis and the SSC modelling can be found in \citet{JBG_Mrk501_bump}}
\label{FigureMrk501SEDMagnification}
}
\end{figure*}

Interim results of the numerical procedure are shown in Figs. \ref{FigureNElectrons}, \ref{FigurePPAndEscape}, and \ref{FigureDotnAndnPhotons} and are discussed now.

The results of our iterative determination of $N(\gamma)$ are shown in Fig.~\ref{FigureNElectrons}. The initial function of $N$ was chosen to be
\begin{equation}
N_{\rm{init}}(\gamma) = \frac{2 \cdot \dot N_{\rm i}(\gamma)}{\int ^{\gamma_{{\mathrm{i}},\,1}}_{\gamma'_{\rm{IC,\,min}}(\gamma_{{\mathrm{i}},\,1},x_0)} C(\gamma_{{\mathrm{i}},\,1},\gamma') \, \mathrm{d}\gamma' + \int ^{\gamma_{{\mathrm{i}},\,0}}_{\gamma'_{\rm{IC,\,min}}(\gamma_{{\mathrm{i}},\,0},x_0)} C(\gamma_{{\mathrm{i}},\,0},\gamma') \, \mathrm{d}\gamma'},
\end{equation}
cf.\ Eq.~\ref{EquationRateOfChangeElectrons}. This choice is a compromise between a function that can be evaluated fast and is accurate. It is a good approximation for the final $N$ in the highest energetic regime because the electron spectral production rate is negligible here. For decreasing $\gamma$, a higher number of iteration steps is necessary because the energy transfer in one IC scattering event decreases. We choose the convergence criterion as follows: One iteration is considered as finished, as soon as the relative change $\left| N_{k,\,j-1} - N_{k,\,j} \right| / N_{k,\,j}$ between two successive values is smaller than $0.001$ at the points $\gamma_k > 10 \, x_{0,\,1}^{-1}$ and smaller than $0.001 \cdot (\gamma_k x_{0,\,1} / 10)^{2}$ at the points $\gamma_k \leq 10 \, x_{0,\,1}^{-1}$. In other words, the demanded accuracy increases quadratically the deeper we iterate into the Thomson regime. There, about $j_{\rm{final}}=100$ iteration steps are necessary to meet the convergence criterion, while only two or three steps are necessary in the KN regime, cf.\ Fig.~\ref{FigureNElectrons}. The shallow ridge of $N$ at $\gamma \approx 2 \, x_{0,\,1}^{-1}$ is caused by PP of highly energetic photons with hydrogen Lyman photons (mainly hydrogen Lyman-$\alpha$ photons), while the very shallow ridge at $\gamma \approx 0.5 \, x_{0,\,1}^{-1}$ is caused by PP of highly energetic photons with the helium II Lyman-$\alpha$ photons. In other words, these ridges in $N$ are the effect of peaks of the spectral PP rate (fourth summand in Eq.~\ref{EquationRateOfChangeElectrons}). In the Thomson regime, with decreasing $\gamma$ the course of $N$ deviates more and more from the $\sim \gamma^{-2}$ dependence of the saturated cascade case considered by \citet{Zdz} because particles escape in our scenario. The finite value of the escape time prevents the electrons from penetrating deep into the Thomson regime.

From Fig.~\ref{FigurePPAndEscape} it is obvious that the photon spectral loss rate is dominated by pair absorption in the approximate range\\ $3\,x_{0,\,1}^{-1}<x_\gamma<800\,x_{0,\,1}^{-1}$  corresponding to a photon energy between $19 \, \rm{GeV}$ and $5.1 \, \rm{TeV}$, and by escape outside this energy range. The maximum of the photon spectral loss rate occurs at $x_\gamma \approx 12 \, x_{0,\,1}^{-1}$, corresponding to the maximum of the hydrogen Lyman-$\alpha$ absorption. The optical depths, which correspond to the respective spectral loss rate contributions, are shown in Fig.~\ref{FigurePPAndEscape} on the right-hand side ordinate and were determined by multiplication of the respective spectral loss rate contribution with the escape time $T_{\rm{esc}}$.

We show the photon spectral production rate $\dot n_{\gamma, \, \rm{IC}}$ in Fig.~\ref{FigureDotnAndnPhotons}. From $N$, it inherits the peak around $500 \, x_{0,\,1}^{-1}$ as well as the deviation from a power law in the Thomson regime due to escape \citep[cf.][]{Zdz}. In this figure we also show the spectral number density $n_\gamma$ of highly energetic photons. The contribution $n_{\gamma,\,\mathrm{i}}$ vanishes because of the choice Eq.~\ref{EquationMrk501DistributionHEPs}. The contribution $n_{\gamma,\,\mathrm{IC}}$ is non-vanishing and ensues as a consequence of IC scattering, pair absorption, and escape. The pronounced trough above $x_{\gamma} \approx 4 \, x_{0,\,1}^{-1}$ is due to absorption by PP with photons from the hydrogen Lyman-$\alpha$ line\footnote{Because of the smallness of $K_{{\rm{line}},\,2}$ and $K_{{\rm{line}},\,3}$, absorption features of the hydrogen Lyman-$\beta$ line and series are almost invisible.} (cf.\ Fig.~\ref{FigurePPAndEscape}). The shallow dip of the spectral number density above $x_{\gamma} \approx x_{0,\,1}^{-1}$ is due to absorption on the helium II Lyman-$\alpha$ line. Thus, the slight bump at $x_{\gamma} \approx x_{0,\,1}^{-1}$ results from cascaded photons that are immediately below the threshold for PP and hence are not absorbed. The peaky feature around $500 \, x_{0,\,1}^{-1}$ is more pronounced in $n_{\gamma}$ than in $\dot n_{\gamma, \, \rm{IC}}$ because the spectral loss rate (i.e.\ the denominator of $n_{\gamma,\,\rm{IC}}$) is higher below $500 \, x_{0,\,1}^{-1}$ than at $500 \, x_{0,\,1}^{-1}$. If $n_0$ were continuous and distributed along a wider range of $x$ instead of consisting of a few strong lines, the peak around $500 \, x_{0,\,1}^{-1}$ would be wider and extend to lower energies.

The total energy density of highly energetic photons is obtained by integrating of the spectral energy density $x_{\gamma} \, n_{\gamma}(x_{\gamma})$:
\begin{equation}
\int _0^{x_{\gamma,\,\mathrm{max}}(\gamma_{{\mathrm{i}},\,0},x_{0,\,1})} x_{\gamma} \, n_{\gamma}(x_{\gamma}) \, \mathrm{d}x_{\gamma} \cdot m_{\rm e} c^2 \approx 4.5 \, \mathrm{J}/\mathrm{m}^3
\end{equation}
The total energy density of the low-energy photons is $K_{\rm{lines}} \cdot \sum _{i=1}^4 K_{{\rm{line}},\,i} \cdot m_{\rm e} c^2 \approx 6.5 \, \mathrm{J}/\mathrm{m}^3$ and thus of similar magnitude. Our cascade is therefore indeed linear (cf.\ Sect. \ref{SubsectionModelAssumptions}).

The cascaded spectral flux density $F_{\rm{casc}}$ of highly energetic photons adds to the spectral flux density $F_{\rm{SSC}}$ of photons that is produced farther downstream the jet in a common one-zone SSC scenario. Both these components contribute to the observed spectral flux density $F$. We fit both components, $F_{\rm{SSC}}$ and $F_{\rm{casc}}$, to the multi-wavelength data of Mrk~501 of the night of MJD 56857.98 as presented by the \citet{JBG_Mrk501_bump}. As shown in Fig.~10 of \citet{JBG_Mrk501_bump} and in Fig.~\ref{FigureMrk501SEDMagnification} in the present paper, we can describe the narrow SED feature without the need of extreme SSC parameters. The parameters found by the fit are given for the cascaded emission component in Table \ref{TableMrk501Parameters} of the present paper and for the SSC emission component in Table~6 and Fig.~7 in \citet{JBG_Mrk501_bump}. The emission from a cascade initiated by electrons originating from a vacuum gap can thus explain the observed narrow peak-like feature.

\subsection{Physical inferences}\label{SubsectionPhysicalInferences}

In this subsection we discuss the results of modelling Mrk~501. We assess the findings in a broader context.

\subsubsection{Number and size of emission-line clouds}\label{SubsubsectionClouds}

We compare the assumed hydrogen Lyman-$\alpha$ photon abundance with the hydrogen Lyman-$\alpha$ luminosity of $L_{\rm{Ly} \,\alpha,\,\rm{obs}} = 5.2 \cdot 10^{33} \, \rm{W}$ that was detected from Mrk~501 by \citet{Stocke}. The total number density of all low-energy photons is given by\\$K_{\rm{lines}} \cdot \sum _{i=1}^4 K_{{\rm{line}},\,i} / x_{0,\,i} \approx 3.2 \cdot 10^{18} \, \rm{m}^{-3}$. This number density describes the photon field of the ionised cloud assumed to be interacting with the electron beam from the gap. The total number density of hydrogen Lyman-$\alpha$ photons is determined by $K_{\rm{lines}} \cdot K_{{\rm{line}},\,4} / x_{0,\,4} \approx 2.6 \cdot 10^{18} \, \rm{m}^{-3}$. We further assume that there are $N_{\rm{cl}}$ clouds of approximate radial size $R_{\rm{cl}}$ in the BLR\@. The hydrogen Lyman-$\alpha$ luminosity $L_{\rm{Ly} \,\alpha,\,\rm{cl}}$ of all these clouds is responsible for the observed hydrogen Lyman-$\alpha$ luminosity. Hence, we can set
\begin{equation}
L_{\rm{Ly} \,\alpha,\,\rm{obs}} \overset{!}{=} L_{\rm{Ly} \,\alpha,\,\rm{cl}} = N_{\rm{cl}} \cdot 4 \pi R_{\rm cl}^2 \: K_{\rm{lines}} \frac{K_{{\rm{line}},\,4}}{x_{0,\,4}} \, c \: x_{0,\,4} \, m_{\rm e} c^2.
\label{EquationMrk501LymanalphaLuminosity}
\end{equation}
As BL Lac objects are considered to be evolved objects running short of cold gas supplies \citep{Liodakis,BlazarReview}, we can assume that $N_{\rm{cl}}$ is quite small. On the other hand, to obtain a Gaussian line profile for the emission lines resulting from the Doppler broadening, the number of clouds must not be too small. Then, solving Eq.~\ref{EquationMrk501LymanalphaLuminosity} for the radial size yields an estimate of $R_{\rm cl}$. Choosing $N_{\rm{cl}} = 10$ yields $R_{\rm cl} \approx 1.8 \cdot 10^{11} \, {\rm m}$, which roughly agrees with our value used for the radial size of the interaction region $R$ (cf.\ Table \ref{TableMrk501Parameters}). This size of the cloud corresponds to the size of a red giant, which is suggestive of cloud formation by photospheric abrasion through beam interaction.
In terms of Schwarzschild radii $r_{\rm S}$ for a central BH of mass $M = 10^9 \, M_{\odot}$, it is $R_{\rm cl} = 0.061 \, r_{\rm S}$. 
From the modelling we infer that in Mrk~501 only few major gas clouds are present in the BLR\@. The encounter of a major cloud with electrons from the gap-beam must therefore be a rare event. Presuming that the gas clouds revolve around the BH on unperturbed Keplerian orbits, an encounter of a cloud with the beam of electrons will at best repeat with the Keplerian orbital period of the cloud. These conclusions are in line with the fact that the peak-like SED feature has not been found elsewhere in Mrk~501 data.

\subsubsection{Reprocessing of the accretion flow luminosity}\label{SubsubsectionReprocessing}

The total line luminosity of these $N_{\rm{cl}}$ clouds of radius $R_{\rm cl}$ is determined as
\begin{equation}
L_{\rm{lines,\,tot}} = N_{\rm{cl}} \cdot 4 \pi R_{\rm cl}^2 \: K_{\rm{lines}} \cdot \sum _{i=1}^4 K_{{\rm{line}},\,i} \, c \, m_{\rm e} c^2.
\label{EquationMrk501TotalLuminosity}
\end{equation}
For the above chosen $N_{\rm{cl}} = 10$ and resulting $R_{\rm cl} = 1.8 \cdot 10^{11} \, {\rm m}$, it is $L_{\rm{lines,\,tot}} \approx 7.8 \cdot 10^{33} \, \rm{W}$, a factor 5 lower than the upper limit of the Mrk~501 BLR luminosity found by \citet{1997MNRAS.286..415C}. We recall that this finding is obtained from our modelling and from the observed hydrogen Lyman-$\alpha$ luminosity. In canonical BLR models, the BLR luminosity is assumed to be about $10 \, \%$ of the accretion flow luminosity \citep{FermiBlazarSequence}. As Mrk~501 is a gas-depleted object, its BLR reprocesses even less of the incident accretion flow radiation. Thus, the reprocessing fraction $\xi = L_{\rm{lines,\,tot}}/L_{\rm{ADAF,\,tot}}$ might be lower than $10 \, \%$, say $1 \, \%$. Here, $L_{\rm{ADAF,\,tot}}$ is the total luminosity of the ADAF\@. To compare the emission-line luminosity with the accretion flow luminosity, we apply the synthetic ADAF spectrum brought forth by \citet{Mahadevan}. There, the spectral luminosity of an ADAF as a function of frequency was parametrised by the viscosity parameter $\alpha$, the pressure ratio $\beta$, the inner boundary $R_{\rm in}$ of the hot accretion flow, its outer boundary $R_{\rm out}$, the central BH mass $M$, the electrons' temperature $T_{\rm e}$, and by the dimensionless mass accretion rate $\dot m$, which is defined as the ratio of the mass accretion rate to the Eddington accretion rate. We integrate numerically, and for the purpose of cross checking also analytically, along the photon energy $\epsilon$ over the spectral luminosity $L_{\rm{ADAF}}(\epsilon)$ \footnote{We integrate essentially over the superposition of the cyclosynchrotron emission, its Comptonised part, and of the bremsstrahlung contribution.} to obtain $L_{\rm{ADAF,\,tot}}$. We use the canonical choices $\alpha = 0.3$, $\beta = 0.5$, $R_{\rm in} = 3 \, r_{\rm S}$, and $R_{\rm out} = 1000 \, r_{\rm S}$ as well as $M = 10^9 \, M_{\odot}$. Furthermore, we choose the energy $\epsilon_{\rm {ADAF},\,0} = 300 \, k_{\rm B} \, T_{\rm e}$ as an upper integration border because the spectrum essentially vanishes above this energy. We let $\dot m$ and $T_{\rm e}$ be free parameters. The chosen requirement $L_{\rm{lines,\,tot}} \approx 0.01 \cdot L_{\rm{ADAF,\,tot}}(\dot m, T_{\rm e})$ can be met with the exemplary $(\dot m, T_{\rm e})$ pairs listed in row 1 of Table \ref{TableMrk501ADAFNumbers}. This shows that the present scenario fits to the general picture of accretion flows as illuminators of BLRs. Moreover, it gives evidence that Mrk~501 is indeed fed by a low-rate, hot ADAF\@. When the reprocessing fraction $\xi$ is increased from $1 \, \%$ to the canonical $10 \, \%$, the values of $T_{\rm e}$ decrease by about $25 \, \%$ (cf.\ Table \ref{TableMrk501ADAFNumbers}).

\subsubsection{Constraints on the in-gap and post-gap multiplication}\label{SubsubsectionPostGap}

The parameter $K_{\rm{G}}$ expresses the number density of electrons that intrude into the interaction region per time interval. In our model, these electrons enter the interaction region after their number was increased in the gap and during the post-gap cascade. The origin of the electron beam along which the multiplication takes place is the vacuum gap (cf.\ Fig.~\ref{FigureGeometrySketch2}), where in turn the origin of the seed electrons is PP through self-interaction of ADAF radiation. In other words, photons that are emitted by the ADAF can collide in the BH magnetosphere and pair-produce seed electrons. We can determine the number density of ADAF photons capable of PP with the help of the above used ADAF model by \citet{Mahadevan}. The spectral number density of ADAF photons is given by $n_{\rm{ADAF}}(\epsilon) = L_{\rm{ADAF}}(\epsilon) / (c \cdot \epsilon \cdot A_{\rm{ADAF}})$, where $A_{\rm{ADAF}} = 2 \, \pi \, R_{\rm out}^2 + 2 \, \pi \, R_{\rm in}^2 + 2 \, \pi \cdot \left( R_{\rm out} \, \sqrt(1.25 \, R_{\rm out}^2) - R_{\rm in} \, \sqrt(1.25 \, R_{\rm in}^2) \right)$ is a rough estimate of the surface area\footnote{We determine the total surface area as a sum of the outer cylindrical lateral area, the inner cylindrical lateral area, and twice the outer conical lateral area subtracted by the inner conical lateral area. Additionally, we assume the height of the ADAF to be $r$ at radius $r$.} of the ADAF\@. As mentioned above, the ADAF spectrum is non-vanishing up to the energy $\epsilon_{\rm {ADAF},\,0}$. Because PP is a threshold process, the lowest possible ADAF photons capable of PP have the energy $\epsilon_{\rm {ADAF,\,PP\,th}} = m_{\rm e}^2 c^4 /\epsilon_{\rm {ADAF},\,0}$. Consequently, the total number density of ADAF photons capable of PP is obtained by
\begin{equation}
n_{{\rm PP}, \, 1} \approx \int _{\epsilon_{\rm {ADAF,\,PP\,th}}}^{\epsilon_{\rm {ADAF},\,0}} n_{\rm{ADAF}}(\epsilon) \, \mathrm{d}\epsilon.
\label{EquationMrk501ADAFTotalNumberDensityPP}
\end{equation}
As a comparison, we determine the total number density of pair-producing MeV photons based on the approximation
\begin{equation}
n_{{\rm PP}, \, 2} \approx 1.4 \cdot 10^{17} \, \dot m^2 \,\frac{10^9 M_{\odot}}{M} \, \rm{m}^{-3}
\label{EquationLRMrk501ADAFTotalNumberDensity}
\end{equation}
from \citet{LR}, which is motivated by findings for the bremsstrahlung luminosity of an ADAF by \citet{NY}. For the ADAF parameters we used above, the assumed $\xi = 1 \, \%$, and the constrained pairs of $\dot m$ and $T_{\rm e}$, we give the correspondingly values of $n_{{\rm PP}, \, 1}$ and $n_{{\rm PP}, \, 2}$ in the second and third row of Table \ref{TableMrk501ADAFNumbers}. Except for the last pair, we have an order-of-magnitude agreement. In contrast, for the canonical $\xi = 10 \, \%$ and the corresponding lower $T_{\rm e}$, we obtain values for $n_{{\rm PP}, \, 1}$ that are about two orders of magnitude lower (cf.\ Table \ref{TableMrk501ADAFNumbers}) than in the $1 \, \%$ reprocessing fraction case.

From the number density of pair-producing ADAF photons, we estimate the materialisation rate in the gap. With $0.2 \, \sigma_{\rm Th}$ as an approximation for the PP cross section, the number density of pairs that are produced in the gap region per unit time is $K_{\rm{PP,\,gap}} \approx 0.2 \, \sigma_{\rm Th} \, n_{{\rm PP}, \, 1}^2 \, c$. We show the values of $K_{\rm{PP,\,gap}}$ for the $(\dot m, T_{\rm e})$ pairs in the fourth row of Table \ref{TableMrk501ADAFNumbers}. All four estimates in the $\xi = 1 \, \%$ case are about six orders of magnitude smaller than our determined $K_{\rm{G}}$ and all four estimates in the $\xi = 10 \, \%$ case are about ten orders smaller. This does not necessarily cast doubt on our model.

These pair-produced seed electrons are accelerated in the gap and multiplied in the gap and during the propagation to the interaction with the emission-line photons. The multiplication is mediated by the emission of synchro-curvature radiation in the magnetospheric magnetic field, by the emission of IC up-scattered radiation in the ADAF photon field, and by the subsequent materialisation of this radiation through PP\@. In net effect, this multiplication increases the number of electrons by the number $\mathcal{N}$ and also lowers the energy per electron by $\mathcal{N}$. Because we neither include the magnetic field and the emission of synchro-curvature radiation nor the exact spatial distribution of the ADAF photon field into our model, a detailed analysis of the in-gap and post-gap multiplication is beyond the scope of this work. We can, however, estimate from comparison of $K_{\rm{PP,\,gap}}$ with $K_{\rm{G}}$ that it has to be $\mathcal{N} \approx 10^6$ in the $1 \, \%$ reprocessing fraction case and $\mathcal{N} \approx 10^{10}$ in the $10 \, \%$ case. The finding $\mathcal{N} \approx 10^6$ is three orders of magnitude larger than the multiplication found by \citet{Broderick}. These authors, however, determined the multiplication only in the post-gap cascade. Considering that up to $10^7$ curvature photons can be emitted by one electron in the initial spark event of the gap \citep{LevinsonCerutti}, $\mathcal{N} \approx 10^6$ after the post-gap cascade is plausible and does not inhibit gap stability. Furthermore, to assess the magnitude of $\mathcal{N}$, we have to consider the magnetic field structure (especially the divergence of the magnetic field), which can dilute the number density of the electron bunch.

The maximum potential drop in the gap is connected with the maximum Lorentz factor of the electrons. Knowing from our modelling (cf.\ Eq.~\ref{EquationMrk501DistributionElectrons} and Table \ref{TableMrk501Parameters}) that the maximum energy per electron after the multiplication is $\gamma_{{\rm{i}},\,0} \, m_{\rm{e}} c^2 = (\gamma_{\rm{mean}} + 3 \, \varsigma) \, m_{\rm{e}} c^2 = 5.7 \cdot\nobreak 10^{12} \, \rm{eV}$, we can estimate that the maximum energy that one original seed electron has attained in the gap voltage is\\ $\approx \mathcal{N} \cdot \gamma_{{\rm{i}},\,0} \, m_{\rm{e}} c^2$. In the $\xi = 1 \, \%$ case, this gives $\approx 5.7 \cdot 10^{18} \, \rm{eV}$, which agrees with estimated \citep{LR} and simulated \citep{HirotaniPu1} values for the potential drop in a vacuum gap. In the $\xi = 10 \, \%$ case, this gives $\approx 5.7 \cdot 10^{22} \, \rm{eV}$, which seems hard to reach even for extreme magnetospheric gap conditions. Thus, the canonical choice $\xi = 10 \, \%$ for the reprocessing fraction does not fit into a consistent picture of Mrk~501. \citet{Levinson}, \citet{LevinsonCerutti}, \citet{Chen2}, \citet{Crinquand}, and \citet{2020arXiv200702838K} have recently cast doubt on the prior attitude of vacuum gaps being steady-state phenomena. Instead, they might be of intermittent or cyclic character. Concretely, \citet{LevinsonCerutti} found that charge-starved gaps, after ignition by injection of gamma rays or electrons abruptly discharge in a gamma-ray flaring event on a timescale $r_{\rm S}/(2 c)$, followed by relaxation towards a state of weak-amplitude oscillations of the electric field and plasma. In this case, rule-of-thumb estimates for the maximum potential drop would not be adequate for conclusions about the attained electron Lorentz factor. We therefore leave detailed inferences from our input parameters about conditions in the vacuum gap to future studies.

If the electron beam is emitted from an oscillating gap, one might expect that this also leads to cyclic behaviour of the cascade with the emission-line photons. However, \citet{Broderick} found that if the gap width exceeds its height, then the gap consists of a number of causally independent tubes. The cyclic behaviour after the initial spark can therefore hardly be traced because the emitted electron beam is the transverse average of many independently fluctuating gap tubes.

\begin{table}
\parbox{.555\linewidth}{
\begin{tabular}{| @{\hspace{0.3cm}}c@{\hspace{0.3cm}} | r@{\hspace{0.4cm}} r@{.}l | c | @{\hspace{0.3cm}}c@{\hspace{0.3cm}} |} 
\hline
\multirow{2}{*}{$i$} 	& \multicolumn{3}{|c|}{Wavelength}								& Relative flux density 							& \multirow{2}{*}{Line designation} \\
											& \multicolumn{3}{|c|}{$\lambda_{0,\,i} / \rm{nm}$} & contribution $K_{{\rm{line}},\,i}$	& \\
\hline
1		& & 30&5 & 2.00	& Helium II Lyman-$\alpha$	\\
2		& & 93&0 & 0.17	& Hydrogen Lyman series			\\
3		& & 102&6 & 0.57	& Hydrogen Lyman-$\beta$		\\
4		& & 121&5 & 5.40	& Hydrogen Lyman-$\alpha$		\\
\hline
\end{tabular}
\caption{Four emission lines used as external low-energy target photons.}
\label{TableMrk501Lines}\vspace{-0.5cm}
\tablefoot{$K_{\rm{line},\,3}$ and $K_{\rm{line},\,4}$ are the relative flux density contributions found by \citet{Pian}. $K_{\rm{line},\,1}$ and $K_{\rm{line},\,2}$ were adopted from \citet{Abolmasov}.}
}
\hspace{1.6cm}
\parbox{.285\linewidth}{
\begin{tabular}{| l@{\hspace{0.4cm}} | l@{\hspace{0.4cm}} |} 
\hline
Quantity									& Used value																							\\
\hline
$\phi$										& $1.8 \degr$																					\\
$R$												& $3.0 \cdot 10^{11} \, \rm{m}$														\\
$K_{\rm{G}}$							& $3.3 \cdot 10^4 \, \rm{s^{-1} m^{-3}}$ 									\\
$K_{\rm{lines}}$					& $9.7 \cdot 10^{12} \, \rm{m^{-3}}$ 											\\
$\gamma_{\mathrm{mean}}$	& $3.4 \cdot 10^{12} \, {\rm{eV}} / (m_{\rm{e}} c^2)$	\\
$\varsigma$									& $0.23 \, \gamma_{\mathrm{mean}}$												\\
\hline
\end{tabular}
\caption{Fitting parameters that were used to describe the narrow SED feature by emission from an IC pair cascade.}
\label{TableMrk501Parameters}
}\hfill
\end{table}

\subsubsection{Comparison of energy budgets}\label{SubsubsectionEnergyBudgets}

The total power that is injected into the interaction volume is determined to be $P_{\rm i} = 4 \pi R^3/3 \cdot \int _{\gamma_{{\mathrm{i}},\,1}} ^{\gamma_{{\mathrm{i}},\,0}} \gamma \dot N_{\rm i}(\gamma) {\rm d}\gamma \cdot m_{\rm e} c^2 = 1.9 \cdot 10^{33} \, {\rm W}$, where $R$ is the radial size of the interaction region. This power is due to the electrons that have inherited their energy from gap acceleration. Hence, $P_{\rm i}$ has to be smaller than the maximum power $P_{\rm gap}$ that can be extracted from the gap, which itself cannot exceed the corresponding Blandford-Znajek power $P_{\rm BZ}$. Moreover, we determine the power, that is radiated by the electrons in the SSC emitting zone to be $P_{\rm SSC} = 4.4 \cdot 10^{36} \, {\rm W}$.

The magnetic flux density at the BH horizon upon assuming an open gap was estimated by \citet{AharonianVariability} to be $B_{\rm BH} = 0.19 \cdot (\beta_{\rm m}/M_9)^{4/7} \, {\rm T}$, where $\beta_{\rm m}$ is the magnetisation of the ADAF\@. Assuming the maximum possible electric field strength, gap extension, BH spin, and charge density, \citet{AharonianVariability} obtained $P_{\rm gap} = 4.8 \cdot 10^{37} \, \beta_{\rm m}^{8/7} \, \kappa \, h \, M_9^{6/7} \ \sin^2\theta \,\, {\rm W}$, where we have substituted for the variability timescale by the height $h$ of the gap in units of Schwarzschild radii, and where $\theta$ is the polar angle, and $\kappa$ is the multiplicity in the gap as defined by \citet{AharonianVariability}. For the values $\theta=\pi/4$, $\kappa=0.1$ (weakly screened gap), $h=0.25$ (thickness is half a gravitational radius), and $\beta_{\rm m}=0.5$ (magnetic pressure is half the gas pressure), we find $P_{\rm gap}=2.7 \cdot 10^{35} \, {\rm W}$. For the same parameters and for the dimensionless BH spin parameter $a_{\ast} = 0.99$, we find $P_{\rm BZ}= 10^{38} \, a_{\ast}^2 \, M_9 \, (B_{\rm BH}/{\rm T})^2 \, {\rm W}= 1.6 \cdot 10^{36} \, {\rm W}$ \citep{HirotaniPu2}, which is twice the accretion flow luminosity (cf.\ Sect. \ref{SubsubsectionReprocessing}.

To conclude, we find $P_{\rm i} / P_{\rm BZ} \approx 0.001$. This agrees with estimates by \citet{Chen2} and \citet{2020arXiv200702838K} and especially with the finding of \citet{LevinsonCerutti}, that a fraction $\approx 0.001$ of $P_{\rm BZ}$ is dissipated in a time span $\approx 7 \, r_{\rm S}/(2 \, c)$ after the initial discharge of a gap. When we again assume $M_9 = 1$, this time span is $\approx 9.6 \, {\rm h}$, the approximate duration of the night of observation of the narrow feature. The comparison $P_{\rm gap} / P_{\rm BZ} \approx 0.2$ is due to the upper limit nature of the estimates by \citet{AharonianVariability}. The ratio $P_{\rm SSC} / P_{\rm BZ}\approx 3$ is acceptable within the uncertainty of $M$ and changes of the magnetic field strength.

\begin{table}
\begin{tabular}{| c | l | l | l | l |}
\hline
\multirow{2}{*}{$(\dot m, T_{\rm e})$} & $(1.0 \cdot 10^{-4}, 1.7 \cdot 10^{10} \, \rm{K})$ & $(2.0 \cdot 10^{-4}, 1.4 \cdot 10^{10} \, \rm{K})$ & $(4.0 \cdot 10^{-4}, 1.1 \cdot 10^{10} \, \rm{K})$ & $(8.0 \cdot 10^{-4}, 8.4 \cdot 10^{9} \, \rm{K})$ \\
& $(1.0 \cdot 10^{-4}, 1.3 \cdot 10^{10} \, \rm{K})$ & $(2.0 \cdot 10^{-4}, 1.0 \cdot 10^{10} \, \rm{K})$ & $(4.0 \cdot 10^{-4}, 8.1 \cdot 10^{9} \, \rm{K})$ & $(8.0 \cdot 10^{-4}, 6.4 \cdot 10^{9} \, \rm{K})$\\
\hline
\multirow{2}{*}{$n_{{\rm PP}, \, 1}/{\rm m}^{-3}$} & $2.8 \cdot 10^9$ & $4.6 \cdot 10^9$ & $4.7 \cdot 10^9$ & $3.8 \cdot 10^9$ \\
& $6.7 \cdot 10^7$ & $3.8 \cdot 10^7$ & $5.5 \cdot 10^7$ & $6.9 \cdot 10^7$ \\
\hline
 $n_{{\rm PP}, \, 2}/{\rm m}^{-3}$ & $1.4 \cdot 10^9$ & $5.6 \cdot 10^9$ & $2.2 \cdot 10^{10}$ & $9.0 \cdot 10^{10}$ \\
\hline
\multirow{2}{*}{$K_{\rm{PP,\,gap}}/\rm{s^{-1} m^{-3}}$} & 0.031 & 0.084 & 0.088 & 0.058 \\
& $1.8 \cdot 10^{-5}$ & $5.8 \cdot 10^{-6}$ & $1.2 \cdot 10^{-5}$ & $1.9 \cdot 10^{-5}$ \\
\hline
\end{tabular}
\caption{Four combinations of physical quantities connecting ADAF properties to the materialisation rate in the vacuum gap. The first row shows pairs of values for the dimensionless accretion rate and for the electron temperature of an ADAF that are permitted by reconciliation of the ADAF luminosity with the luminosity of the line emitting clouds. The corresponding values of the total number density of pair producing ADAF photons according to the model by \citet{Mahadevan}, of the total number density of pair producing ADAF photons according to the estimation by \citet{LR}, and of the materialisation rate of pair-produced electrons in the gap region are shown in the second, third, and fourth line, respectively. In each field, the upper item was obtained with the assumption of $\xi = 1 \, \%$ and the lower item was obtained with $\xi = 10 \, \%$. As $n_{{\rm PP}, \, 2}$ is independent of $T_{\rm e}$, it has the same value for both cases.}
\label{TableMrk501ADAFNumbers}
\end{table}

\section{Summary}\label{SectionSummary}

We have modelled the interaction of particle beams from vacuum gaps in the magnetospheres of spinning BHs in blazars with recombination-line photons from surrounding gas clouds. Such clouds are commonly found in the BLR of quasars, but could also arise in early-type galaxies as a result of red giants that cross the particle beam. Numerically solving the coupled kinetic equations corresponding to linear IC pair cascades with escape terms, we obtained the steady-state solution of the SED of the photons escaping from the interaction zone. Compared to previous studies of cascades in AGN, our approach considered three main ramifications:
\begin{itemize}
	\item The inclusion of an escape term corresponds to a source that is finitely extended and not necessarily optically thick. Thus, in contrast to the work of \citet{Zdz}, which dealt with saturated and hence optically thick cascades, our scenario can be applied to astrophysical objects that are optically thin to the emission of HE gamma rays. This is only a first step towards the self-consistent treatment of a spherically symmetric emission region with a radially varying emission and absorption coefficients. For the SSC mechanism without pair cascades, such a treatment was developed by \citet{Gould}. In our framework, the photon density reacts back upon itself through the electron density. Hence, we did not treat the emission coefficient as prescribed.
	\item \citet{Wendel17} determined the photon spectral number density $n_\gamma$ essentially as the accumulation of all the IC up-scattered photons. In contrast, we determined $n_\gamma$ by considering both IC up-scattering and losses through escape and pair absorption (cf.\ Eq.~\ref{EquationSpectralNumberDensityHEPs}).
	\item We used an efficient numerical iteration scheme (cf.\ Sect. \ref{SectionNumericalSolutionProcedure}) that saves computational cost by iterating from values $\gamma_k$ with high $k$ to lower $k$, and so using the optimal course of $N(\gamma)$ in each iteration step.
\end{itemize}
Applying the model to the blazar Mrk~501, we showed that the recently observed peculiar narrow spectral feature at $3$~TeV can readily be explained when rather plausible physical parameters for the beam and recombination-line clouds are adopted. In the scenario supported by this observation, photons from a weakly accreting hot ADAF materialise as pairs in the central BH magnetosphere. These seed electrons are accelerated in a vacuum gap and are multiplied by post-gap cascades while travelling away from the central region and carry about $0.1 \, \%$ of the Blandford-Znajek luminosity. The ADAF also irradiates and ionises ambient gas clouds. In agreement with measurements of the hydrogen Lyman-$\alpha$ luminosity in Mrk~501, these clouds reprocess about $1 \, \%$ of the original ADAF luminosity as emission-line photons. They act as a target colliding with the electron beam. The resulting cascade photons are emitted in the direction of the primary beam electrons, shaping the intermittent narrow bump at about 3\,TeV in the SED\@. In combination with the SSC spectrum produced by particles presumably accelerated at shock waves further downstream, the model explains the broadband SED of Mrk~501 at MJD 56857.98, assuming no further pair attenuation within the host galaxy to occur. Predictions about the duty cycle of this feature are difficult considering the intermittency of gap formation and randomness of nearby cloud passages.

\begin{acknowledgements}
We thank Dorit Glawion, and Amit Shukla for fruitful discussions on the topic. We are also thankful to the anonymous referee for constructive criticism on the manuscript. Moreover, we thank Astrid Peter and Julian Sitarek for comments on this work as well as J. D. Hunter and co-workers for the development of matplotlib \citep{Hunter}. C. W. gratefully acknowledges support by the project "Promotion inklusive" of the Universit\"{a}t zu K\"{o}ln and the German Bundesministerium f\"{u}r Arbeit und Soziales. J. B. G. acknowledges the support of the Viera y Clavijo program funded by ACIISI and ULL\@.
\end{acknowledgements}

\bibliographystyle{aa} 
\bibliography{Article-AA-2020-38343}

\begin{thebibliography}{68}
\expandafter\ifx\csname natexlab\endcsname\relax\def\natexlab#1{#1}\fi

\bibitem[{{Abdo} {et~al.}(2011){Abdo}, {Ackermann}, {Ajello}, {Allafort},
  {Baldini}, {Ballet}, {Barbiellini}, {Baring}, {Bastieri}, {Bechtol},
  {Bellazzini}, {Berenji}, {Blandford}, {Bloom}, {Bonamente}, {Borgland},
  {Bouvier}, {Brandt}, {Bregeon}, {Brez}, {Brigida}, {Bruel}, {Buehler},
  {Buson}, {Caliandro}, {Cameron}, {Cannon}, {Caraveo}, {Carrigan},
  {Casandjian}, {Cavazzuti}, {Cecchi}, {{\c{C}}elik}, {Charles}, {Chekhtman},
  {Cheung}, {Chiang}, {Ciprini}, {Claus}, {Cohen-Tanugi}, {Conrad}, {Cutini},
  {Dermer}, {de Palma}, {Silva}, {Drell}, {Dubois}, {Dumora}, {Favuzzi},
  {Fegan}, {Ferrara}, {Focke}, {Fortin}, {Frailis}, {Fuhrmann}, {Fukazawa},
  {Funk}, {Fusco}, {Gargano}, {Gasparrini}, {Gehrels}, {Germani}, {Giglietto},
  {Giordano}, {Giroletti}, {Glanzman}, {Godfrey}, {Grenier}, {Guillemot},
  {Guiriec}, {Hayashida}, {Hays}, {Horan}, {Hughes}, {J{\'o}hannesson},
  {Johnson}, {Johnson}, {Kadler}, {Kamae}, {Katagiri}, {Kataoka},
  {Kn{\"o}dlseder}, {Kuss}, {Lande}, {Latronico}, {Lee}, {Lemoine-Goumard},
  {Longo}, {Loparco}, {Lott}, {Lovellette}, {Lubrano}, {Madejski}, {Makeev},
  {Max-Moerbeck}, {Mazziotta}, {McEnery}, {Mehault}, {Michelson},
  {Mitthumsiri}, {Mizuno}, {Moiseev}, {Monte}, {Monzani}, {Morselli},
  {Moskalenko}, {Murgia}, {Naumann-Godo}, {Nishino}, {Nolan}, {Norris}, {Nuss},
  {Ohsugi}, {Okumura}, {Omodei}, {Orlando}, {Ormes}, {Paneque}, {Panetta},
  {Parent}, {Pavlidou}, {Pearson}, {Pelassa}, {Pepe}, {Pesce-Rollins}, {Piron},
  {Porter}, {Rain{\`o}}, {Rando}, {Razzano}, {Readhead}, {Reimer}, {Reimer},
  {Richards}, {Ripken}, {Ritz}, {Roth}, {Sadrozinski}, {Sanchez}, {Sander},
  {Scargle}, {Sgr{\`o}}, {Siskind}, {Smith}, {Spandre}, {Spinelli}, {Stawarz},
  {Stevenson}, {Strickman}, {Sokolovsky}, {Suson}, {Takahashi}, {Takahashi},
  {Tanaka}, {Thayer}, {Thayer}, {Thompson}, {Tibaldo}, {Torres}, {Tosti},
  {Tramacere}, {Uchiyama}, {Usher}, {Vandenbroucke}, {Vasileiou}, {Vilchez},
  {Vitale}, {Waite}, {Wang}, {Wehrle}, {Winer}, {Wood}, {Yang}, {Ylinen},
  {Zensus}, {Ziegler}, {Fermi LAT Collaboration}, {Aleksi{\'c}}, {Antonelli},
  {Antoranz}, {Backes}, {Barrio}, {Becerra Gonz{\'a}lez}, {Bednarek},
  {Berdyugin}, {Berger}, {Bernardini}, {Biland}, {Blanch}, {Bock}, {Boller},
  {Bonnoli}, {Bordas}, {Borla Tridon}, {Bosch-Ramon}, {Bose}, {Braun}, {Bretz},
  {Camara}, {Carmona}, {Carosi}, {Colin}, {Colombo}, {Contreras}, {Cortina},
  {Covino}, {Dazzi}, {de Angelis}, {De Cea del Pozo}, {De Lotto}, {De Maria},
  {De Sabata}, {Delgado Mendez}, {Diago Ortega}, {Doert}, {Dom{\'\i}nguez},
  {Dominis Prester}, {Dorner}, {Doro}, {Elsaesser}, {Ferenc}, {Fonseca},
  {Font}, {Garc{\'\i}a L{\'o}pez}, {Garczarczyk}, {Gaug}, {Giavitto},
  {Godinovi}, {Hadasch}, {Herrero}, {Hildebrand}, {H{\"o}hne-M{\"o}nch},
  {Hose}, {Hrupec}, {Jogler}, {Klepser}, {Kr{\"a}henb{\"u}hl}, {Kranich},
  {Krause}, {La Barbera}, {Leonardo}, {Lindfors}, {Lombardi}, {L{\'o}pez},
  {Lorenz}, {Majumdar}, {Makariev}, {Maneva}, {Mankuzhiyil}, {Mannheim},
  {Maraschi}, {Mariotti}, {Mart{\'\i}nez}, {Mazin}, {Meucci}, {Miranda},
  {Mirzoyan}, {Miyamoto}, {Mold{\'o}n}, {Moralejo}, {Nieto}, {Nilsson},
  {Orito}, {Oya}, {Paoletti}, {Paredes}, {Partini}, {Pasanen}, {Pauss},
  {Pegna}, {Perez-Torres}, {Persic}, {Peruzzo}, {Pochon}, {Prada Moroni},
  {Prada}, {Prandini}, {Puchades}, {Puljak}, {Reichardt}, {Reinthal}, {Rhode},
  {Rib{\'o}}, {Rico}, {Rissi}, {R{\"u}gamer}, {Saggion}, {Saito}, {Saito},
  {Salvati}, {S{\'a}nchez-Conde}, {Satalecka}, {Scalzotto}, {Scapin},
  {Schultz}, {Schweizer}, {Shayduk}, {Shore}, {Sierpowska-Bartosik},
  {Sillanp{\"a}{\"a}}, {Sitarek}, {Sobczynska}, {Spanier}, {Spiro}, {Stamerra},
  {Steinke}, {Storz}, {Strah}, {Struebig}, {Suric}, {Takalo}, {Tavecchio},
  {Temnikov}, {Terzi{\'c}}, {Tescaro}, {Teshima}, {Vankov}, {Wagner},
  {Weitzel}, {Zabalza}, {Zandanel}, {Zanin}, {MAGIC Collaboration}, {Acciari},
  {Arlen}, {Aune}, {Benbow}, {Boltuch}, {Bradbury}, {Buckley}, {Bugaev},
  {Cannon}, {Cesarini}, {Ciupik}, {Cui}, {Dickherber}, {Errando}, {Falcone},
  {Finley}, {Finnegan}, {Fortson}, {Furniss}, {Galante}, {Gall}, {Gillanders},
  {Godambe}, {Grube}, {Guenette}, {Gyuk}, {Hanna}, {Holder}, {Huang}, {Hui},
  {Humensky}, {Kaaret}, {Karlsson}, {Kertzman}, {Kieda}, {Konopelko},
  {Krawczynski}, {Krennrich}, {Lang}, {Maier}, {McArthur}, {McCann},
  {McCutcheon}, {Moriarty}, {Mukherjee}, {Ong}, {Otte}, {Pandel}, {Perkins},
  {Pichel}, {Pohl}, {Quinn}, {Ragan}, {Reyes}, {Reynolds}, {Roache}, {Rose},
  {Rovero}, {Schroedter}, {Sembroski}, {Senturk}, {Steele}, {Swordy},
  {Te{\v{s}}i{\'c}}, {Theiling}, {Thibadeau}, {Varlotta}, {Vincent}, {Wakely},
  {Ward}, {Weekes}, {Weinstein}, {Weisgarber}, {Williams}, {Wood}, {Zitzer},
  {VERITAS Collaboration}, {Villata}, {Raiteri}, {Aller}, {Aller}, {Arkharov},
  {Blinov}, {Calcidese}, {Chen}, {Efimova}, {Kimeridze}, {Konstantinova},
  {Kopatskaya}, {Koptelova}, {Kurtanidze}, {Kurtanidze}, {L{\"a}hteenm{\"a}ki},
  {Larionov}, {Larionova}, {Larionova}, {Ligustri}, {Morozova}, {Nikolashvili},
  {Sigua}, {Troitsky}, {Angelakis}, {Capalbi}, {Carrami{\~n}ana}, {Carrasco},
  {Cassaro}, {de la Fuente}, {Gurwell}, {Kovalev}, {Kovalev}, {Krichbaum},
  {Krimm}, {Leto}, {Lister}, {Maccaferri}, {Moody}, {Mori}, {Nestoras},
  {Orlati}, {Pagani}, {Pace}, {Pearson}, {Perri}, {Piner}, {Pushkarev}, {Ros},
  {Sadun}, {Sakamoto}, {Tornikoski}, {Yatsu}, \& {Zook}}]{2011ApJ...727..129A}
{Abdo}, A.~A., {Ackermann}, M., {Ajello}, M., {et~al.} 2011, \apj, 727, 129

\bibitem[{{Abolmasov} \& {Poutanen}(2017)}]{Abolmasov}
{Abolmasov}, P. \& {Poutanen}, J. 2017, \mnras, 464, 152

\bibitem[{{Agaronyan} {et~al.}(1983){Agaronyan}, {Atoyan}, \&
  {Nagapetyan}}]{Agaronyan}
{Agaronyan}, F.~A., {Atoyan}, A.~M., \& {Nagapetyan}, A.~M. 1983, Astrophysics,
  19, 187

\bibitem[{{Aharonian} {et~al.}(2007){Aharonian}, {Akhperjanian}, {Bazer-Bachi},
  {Behera}, {Beilicke}, {Benbow}, {Berge}, {Bernl{\"o}hr}, {Boisson}, {Bolz},
  {Borrel}, {Boutelier}, {Braun}, {Brion}, {Brown}, {B{\"u}hler},
  {B{\"u}sching}, {Bulik}, {Carrigan}, {Chadwick}, {Clapson}, {Chounet},
  {Coignet}, {Cornils}, {Costamante}, {Degrange}, {Dickinson},
  {Djannati-Ata{\"\i}}, {Domainko}, {Drury}, {Dubus}, {Dyks}, {Egberts},
  {Emmanoulopoulos}, {Espigat}, {Farnier}, {Feinstein}, {Fiasson},
  {F{\"o}rster}, {Fontaine}, {Funk}, {Funk}, {F{\"u}{\ss}ling}, {Gallant},
  {Giebels}, {Glicenstein}, {Gl{\"u}ck}, {Goret}, {Hadjichristidis}, {Hauser},
  {Hauser}, {Heinzelmann}, {Henri}, {Hermann}, {Hinton}, {Hoffmann}, {Hofmann},
  {Holleran}, {Hoppe}, {Horns}, {Jacholkowska}, {de Jager}, {Kendziorra},
  {Kerschhaggl}, {Kh{\'e}lifi}, {Komin}, {Kosack}, {Lamanna}, {Latham}, {Le
  Gallou}, {Lemi{\`e}re}, {Lemoine-Goumard}, {Lenain}, {Lohse}, {Martin},
  {Martineau-Huynh}, {Marcowith}, {Masterson}, {Maurin}, {McComb}, {Moderski},
  {Moulin}, {de Naurois}, {Nedbal}, {Nolan}, {Olive}, {Orford}, {Osborne},
  {Ostrowski}, {Panter}, {Pedaletti}, {Pelletier}, {Petrucci}, {Pita},
  {P{\"u}hlhofer}, {Punch}, {Ranchon}, {Raubenheimer}, {Raue}, {Rayner},
  {Renaud}, {Ripken}, {Rob}, {Rolland}, {Rosier-Lees}, {Rowell}, {Rudak},
  {Ruppel}, {Sahakian}, {Santangelo}, {Saug{\'e}}, {Schlenker}, {Schlickeiser},
  {Schr{\"o}der}, {Schwanke}, {Schwarzburg}, {Schwemmer}, {Shalchi}, {Sol},
  {Spangler}, {Stawarz}, {Steenkamp}, {Stegmann}, {Superina}, {Tam},
  {Tavernet}, {Terrier}, {van Eldik}, {Vasileiadis}, {Venter}, {Vialle},
  {Vincent}, {Vivier}, {V{\"o}lk}, {Volpe}, {Wagner}, {Ward}, \&
  {Zdziarski}}]{ReferenceVariability1}
{Aharonian}, F., {Akhperjanian}, A.~G., {Bazer-Bachi}, A.~R., {et~al.} 2007,
  \apj, 664, L71

\bibitem[{{Aharonian} {et~al.}(2017){Aharonian}, {Barkov}, \&
  {Khangulyan}}]{AharonianVariability}
{Aharonian}, F.~A., {Barkov}, M.~V., \& {Khangulyan}, D. 2017, \apj, 841, 61

\bibitem[{{Aharonian} \& {Plyasheshnikov}(2003)}]{Aharonian}
{Aharonian}, F.~A. \& {Plyasheshnikov}, A.~V. 2003, Astroparticle Physics, 19,
  525

\bibitem[{{Ahnen} {et~al.}(2017){Ahnen}, {Ansoldi}, {Antonelli}, {Antoranz},
  {Babic}, {Banerjee}, {Bangale}, {Barres de Almeida}, {Barrio}, {Becerra
  Gonz{\'a}lez}, {Bednarek}, {Bernardini}, {Berti}, {Biasuzzi}, {Biland},
  {Blanch}, {Bonnefoy}, {Bonnoli}, {Borracci}, {Bretz}, {Buson}, {Carosi},
  {Chatterjee}, {Clavero}, {Colin}, {Colombo}, {Contreras}, {Cortina},
  {Covino}, {Da Vela}, {Dazzi}, {De Angelis}, {De Lotto}, {de O{\~n}a
  Wilhelmi}, {Di Pierro}, {Doert}, {Dom{\'\i}nguez}, {Dominis Prester},
  {Dorner}, {Doro}, {Einecke}, {Eisenacher Glawion}, {Elsaesser},
  {Engelkemeier}, {Fallah Ramazani}, {Fern{\'a}ndez-Barral}, {Fidalgo},
  {Fonseca}, {Font}, {Frantzen}, {Fruck}, {Galindo}, {Garc{\'\i}a L{\'o}pez},
  {Garczarczyk}, {Garrido Terrats}, {Gaug}, {Giammaria}, {Godinovi{\'c}},
  {Gonz{\'a}lez Mu{\~n}oz}, {Gora}, {Guberman}, {Hadasch}, {Hahn}, {Hanabata},
  {Hayashida}, {Herrera}, {Hose}, {Hrupec}, {Hughes}, {Idec}, {Kodani},
  {Konno}, {Kubo}, {Kushida}, {La Barbera}, {Lelas}, {Lindfors}, {Lombardi},
  {Longo}, {L{\'o}pez}, {L{\'o}pez-Coto}, {Majumdar}, {Makariev}, {Mallot},
  {Maneva}, {Manganaro}, {Mannheim}, {Maraschi}, {Marcote}, {Mariotti},
  {Mart{\'\i}nez}, {Mazin}, {Menzel}, {Mirand a}, {Mirzoyan}, {Moralejo},
  {Moretti}, {Nakajima}, {Neustroev}, {Niedzwiecki}, {Nievas Rosillo},
  {Nilsson}, {Nishijima}, {Noda}, {Nogu{\'e}s}, {Overkemping}, {Paiano},
  {Palacio}, {Palatiello}, {Paneque}, {Paoletti}, {Paredes}, {Paredes-Fortuny},
  {Pedaletti}, {Peresano}, {Perri}, {Persic}, {Poutanen}, {Prada Moroni},
  {Prandini}, {Puljak}, {Reichardt}, {Rhode}, {Rib{\'o}}, {Rico}, {Rodriguez
  Garcia}, {Saito}, {Satalecka}, {Schr{\"o}der}, {Schultz}, {Schweizer},
  {Shore}, {Sillanp{\"a}{\"a}}, {Sitarek}, {Snidaric}, {Sobczynska},
  {Stamerra}, {Steinbring}, {Strzys}, {Suri{\'c}}, {Takalo}, {Tavecchio},
  {Temnikov}, {Terzi{\'c}}, {Tescaro}, {Teshima}, {Thaele}, {Torres}, {Toyama},
  {Treves}, {Vanzo}, {Verguilov}, {Vovk}, {Ward}, {Will}, {Wu}, {Zanin},
  {Abeysekara}, {Archambault}, {Archer}, {Benbow}, {Bird}, {Buchovecky},
  {Buckley}, {Bugaev}, {Connolly}, {Cui}, {Dickinson}, {Falcone}, {Feng},
  {Finley}, {Fleischhack}, {Flinders}, {Fortson}, {Gillanders}, {Griffin},
  {Grube}, {H{\"u}tten}, {Hanna}, {Holder}, {Humensky}, {Kaaret}, {Kar},
  {Kelley-Hoskins}, {Kertzman}, {Kieda}, {Krause}, {Krennrich}, {Lang},
  {Maier}, {McCann}, {Moriarty}, {Mukherjee}, {Nieto}, {O'Brien}, {Ong},
  {Otte}, {Park}, {Perkins}, {Pichel}, {Pohl}, {Popkow}, {Pueschel}, {Quinn},
  {Ragan}, {Reynolds}, {Richards}, {Roache}, {Rovero}, {Rulten}, {Sadeh},
  {Santander}, {Sembroski}, {Shahinyan}, {Telezhinsky}, {Tucci}, {Tyler},
  {Wakely}, {Weinstein}, {Wilcox}, {Wilhelm}, {Williams}, {Zitzer}, {Razzaque},
  {Villata}, {Raiteri}, {Aller}, {Aller}, {Larionov}, {Arkharov}, {Blinov},
  {Efimova}, {Grishina}, {Hagen-Thorn}, {Kopatskaya}, {Larionova}, {Larionova},
  {Morozova}, {Troitsky}, {Ligustri}, {Calcidese}, {Berdyugin}, {Kurtanidze},
  {Nikolashvili}, {Kimeridze}, {Sigua}, {Kurtanidze}, {Chigladze}, {Chen},
  {Koptelova}, {Sakamoto}, {Sadun}, {Moody}, {Pace}, {Pearson}, {Yatsu},
  {Mori}, {Carraminyana}, {Carrasco}, {de la Fuente}, {Norris}, {Smith},
  {Wehrle}, {Gurwell}, {Zook}, {Pagani}, {Perri}, {Capalbi}, {Cesarini},
  {Krimm}, {Kovalev}, {Kovalev}, {Ros}, {Pushkarev}, {Lister}, {Sokolovsky},
  {Kadler}, {Piner}, {L{\"a}hteenm{\"a}ki}, {Tornikoski}, {Angelakis},
  {Krichbaum}, {Nestoras}, {Fuhrmann}, {Zensus}, {Cassaro}, {Orlati},
  {Maccaferri}, {Leto}, {Giroletti}, {Richards}, {Max-Moerbeck}, \&
  {Readhead}}]{2017A&A...603A..31A}
{Ahnen}, M.~L., {Ansoldi}, S., {Antonelli}, L.~A., {et~al.} 2017, \aap, 603,
  A31

\bibitem[{{Ahnen} {et~al.}(2018){Ahnen}, {Ansoldi}, {Antonelli}, {Arcaro},
  {Babi{\'c}}, {Banerjee}, {Bangale}, {Barres de Almeida}, {Barrio}, {Becerra
  Gonz{\'a}lez}, {Bednarek}, {Bernardini}, {Berti}, {Bhattacharyya}, {Blanch},
  {Bonnoli}, {Carosi}, {Carosi}, {Chatterjee}, {Colak}, {Colin}, {Colombo},
  {Contreras}, {Cortina}, {Covino}, {Cumani}, {Da Vela}, {Dazzi}, {De Angelis},
  {De Lotto}, {Delfino}, {Delgado}, {Di Pierro}, {Doert}, {Dom{\'\i}nguez},
  {Dominis Prester}, {Doro}, {Eisenacher lawion}, {Engelkemeier}, {Fallah
  Ramazani}, {Fern{\'a}ndez-Barral}, {Fidalgo}, {Fonseca}, {Font}, {Fruck},
  {Galindo}, {Garc{\'\i}a L{\'o}pez}, {Garczarczyk}, {Gaug}, {Giammaria},
  {Godinovi{\'c}}, {Gora}, {Guberman}, {Hadasch}, {Hahn}, {Hassan},
  {Hayashida}, {Herrera}, {Hose}, {Hrupec}, {Ishio}, {Konno}, {Kubo},
  {Kushida}, {Kuve{\v{z}}di{\'c}}, {Lelas}, {Lindfors}, {Lombardi}, {Longo},
  {L{\'o}pez}, {Maggio}, {Majumdar}, {Makariev}, {Maneva}, {Manganaro},
  {Maraschi}, {Mariotti}, {Mart{\'\i}nez}, {Mazin}, {Menzel}, {Minev},
  {Miranda}, {Mirzoyan}, {Moralejo}, {Moreno}, {Moretti}, {Nagayoshi},
  {Neustroev}, {Niedzwiecki}, {Nievas Rosillo}, {Nigro}, {Nilsson}, {Ninci},
  {Nishijima}, {Noda}, {Nogu{\'e}s}, {Paiano}, {Palacio}, {Paneque},
  {Paoletti}, {Paredes}, {Pedaletti}, {Peresano}, {Perri}, {Persic}, {Prada
  Moroni}, {Prand ini}, {Puljak}, {Garcia}, {Reichardt}, {Rib{\'o}}, {Rico},
  {Righi}, {Rugliancich}, {Saito}, {Satalecka}, {Schroeder}, {Schweizer},
  {Shore}, {Sitarek}, {{\v{S}}nidari{\'c}}, {Sobczynska}, {Stamerra}, {Strzys},
  {Suri{\'c}}, {Takalo}, {Tavecchio}, {Temnikov}, {Terzi{\'c}}, {Teshima},
  {Torres-Alb{\`a}}, {Treves}, {Tsujimoto}, {Vanzo}, {Vazquez Acosta}, {Vovk},
  {Ward}, {Will}, {Zari{\'c}}, {Arbet-Engels}, {Baack}, {Balbo}, {Biland},
  {Blank}, {Bretz}, {Bruegge}, {Bulinski}, {Buss}, {Dmytriiev}, {Dorner},
  {Einecke}, {Elsaesser}, {Herbst}, {Hildebrand}, {Kortmann}, {Linhoff},
  {Mahlke}, {Mannheim}, {Mueller}, {Neise}, {Neronov}, {Noethe}, {Oberkirch},
  {Paravac}, {Rhode}, {Schleicher}, {Schulz}, {Sedlaczek}, {Shukla}, {Sliusar},
  {Walter}, {Archer}, {Benbow}, {Bird}, {Brose}, {Buckley}, {Bugaev},
  {Christiansen}, {Cui}, {Daniel}, {Falcone}, {Feng}, {Finley}, {Gilland ers},
  {Gueta}, {Hanna}, {Hervet}, {Holder}, {Hughes}, {H{\"u}tten}, {Humensky},
  {Johnson}, {Kaaret}, {Kar}, {Kelley-Hoskins}, {Kertzman}, {Kieda}, {Krause},
  {Krennrich}, {Kumar}, {Lang}, {Lin}, {Maier}, {McArthur}, {Moriarty},
  {Mukherjee}, {O'Brien}, {Ong}, {Otte}, {Park}, {Petrashyk}, {Pichel}, {Pohl},
  {Quinn}, {Ragan}, {Reynolds}, {Richards}, {Roache}, {Rovero}, {Rulten},
  {Sadeh}, {Santander}, {Sembroski}, {Shahinyan}, {Sushch}, {Tyler}, {Wakely},
  {Weinstein}, {Wells}, {Wilcox}, {Wilhel}, {Williams}, {J Williamson},
  {Zitzer}, {Perri}, {Verrecchia}, {Leto}, {Villata}, {Raiteri}, {Jorstad},
  {Larionov}, {Blinov}, {Grishina}, {Kopatskaya}, {Larionova}, {Nikiforova},
  {Morozova}, {Troitskaya}, {Troitsky}, {Kurtanidze}, {Nikolashvili},
  {Kurtanidze}, {Kimeridze}, {Chigladze}, {Strigachev}, {Sadun}, {Moody},
  {Chen}, {Lin}, {Acosta-Pulido}, {Ar{\'e}valo}, {Carnerero},
  {Gonz{\'a}lez-Morales}, {Manilla-Robles}, {Jermak}, {Steele}, {Mundell},
  {Ben{\'\i}tez}, {Hiriart}, {Smith}, {Max-Moerbeck}, {Readhead}, {Richards},
  {Hovatta}, {L{\"a}hteenm{\"a}ki}, {Tornikoski}, {Tammi}, {Georganopoulos}, \&
  {Baring}}]{Ahnen18}
{Ahnen}, M.~L., {Ansoldi}, S., {Antonelli}, L.~A., {et~al.} 2018, \aap, 620,
  A181

\bibitem[{{Akharonian} {et~al.}(1985){Akharonian}, {Kririllov-Ugriumov}, \&
  {Vardanian}}]{1985Akharonian}
{Akharonian}, F.~A., {Kririllov-Ugriumov}, V.~G., \& {Vardanian}, V.~V. 1985,
  \apss, 115, 201

\bibitem[{{Albert} {et~al.}(2007){Albert}, {Aliu}, {Anderhub}, {Antoranz},
  {Armada}, {Baixeras}, {Barrio}, {Bartko}, {Bastieri}, {Becker}, {Bednarek},
  {Berger}, {Bigongiari}, {Biland}, {Bock}, {Bordas}, {Bosch- Ramon}, {Bretz},
  {Britvitch}, {Camara}, {Carmona}, {Chilingarian}, {Coarasa}, {Commichau},
  {Contreras}, {Cortina}, {Costado}, {Curtef}, {Danielyan}, {Dazzi}, {De
  Angelis}, {Delgado}, {de los Reyes}, {De Lotto}, {Domingo-Santamar{\'\i}a},
  {Dorner}, {Doro}, {Errando}, {Fagiolini}, {Ferenc}, {Fern{\'a}ndez}, {Firpo},
  {Flix}, {Fonseca}, {Font}, {Fuchs}, {Galante}, {Garc{\'\i}a-L{\'o}pez},
  {Garczarczyk}, {Gaug}, {Giller}, {Goebel}, {Hakobyan}, {Hayashida},
  {Hengstebeck}, {Herrero}, {H{\"o}hne}, {Hose}, {Hrupec}, {Hsu}, {Jacon},
  {Jogler}, {Kosyra}, {Kranich}, {Kritzer}, {Laille}, {Lindfors}, {Lombardi},
  {Longo}, {L{\'o}pez}, {L{\'o}pez}, {Lorenz}, {Majumdar}, {Maneva},
  {Mannheim}, {Mansutti}, {Mariotti}, {Mart{\'\i}nez}, {Mazin}, {Merck},
  {Meucci}, {Meyer}, {Miranda}, {Mirzoyan}, {Mizobuchi}, {Moralejo}, {Nieto},
  {Nilsson}, {Ninkovic}, {O{\~n}a-Wilhelmi}, {Otte}, {Oya}, {Paneque},
  {Panniello}, {Paoletti}, {Paredes}, {Pasanen}, {Pascoli}, {Pauss}, {Pegna},
  {Persic}, {Peruzzo}, {Piccioli}, {Prandini}, {Puchades}, {Raymers}, {Rhode},
  {Rib{\'o}}, {Rico}, {Rissi}, {Robert}, {R{\"u}gamer}, {Saggion}, {Saito},
  {S{\'a}nchez}, {Sartori}, {Scalzotto}, {Scapin}, {Schmitt}, {Schweizer},
  {Shayduk}, {Shinozaki}, {Shore}, {Sidro}, {Sillanp{\"a}{\"a}}, {Sobczynska},
  {Stamerra}, {Stark}, {Takalo}, {Tavecchio}, {Temnikov}, {Tescaro}, {Teshima},
  {Torres}, {Turini}, {Vankov}, {Vitale}, {Wagner}, {Wibig}, {Wittek},
  {Zandanel}, {Zanin}, \& {Zapatero}}]{ReferenceVariability2}
{Albert}, J., {Aliu}, E., {Anderhub}, H., {et~al.} 2007, \apj, 669, 862

\bibitem[{{Aleksi{\'c}} {et~al.}(2014){Aleksi{\'c}}, {Ansoldi}, {Antonelli},
  {Antoranz}, {Babic}, {Bangale}, {Barrio}, {Gonz{\'a}lez}, {Bednarek},
  {Bernardini}, {Biasuzzi}, {Biland}, {Blanch}, {Bonnefoy}, {Bonnoli},
  {Borracci}, {Bretz}, {Carmona}, {Carosi}, {Colin}, {Colombo}, {Contreras},
  {Cortina}, {Covino}, {Da Vela}, {Dazzi}, {De Angelis}, {De Caneva}, {De
  Lotto}, {Wilhelmi}, {Mendez}, {Prester}, {Dorner}, {Doro}, {Einecke},
  {Eisenacher}, {Elsaesser}, {Fonseca}, {Font}, {Frantzen}, {Fruck}, {Galindo},
  {L{\'o}pez}, {Garczarczyk}, {Terrats}, {Gaug}, {Godinovi{\'c}}, {Mu{\~n}oz},
  {Gozzini}, {Hadasch}, {Hanabata}, {Hayashida}, {Herrera}, {Hildebrand},
  {Hose}, {Hrupec}, {Idec}, {Kadenius}, {Kellermann}, {Kodani}, {Konno},
  {Krause}, {Kubo}, {Kushida}, {La Barbera}, {Lelas}, {Lewandowska},
  {Lindfors}, {Lombardi}, {Longo}, {L{\'o}pez}, {L{\'o}pez-Coto},
  {L{\'o}pez-Oramas}, {Lorenz}, {Lozano}, {Makariev}, {Mallot}, {Maneva},
  {Mankuzhiyil}, {Mannheim}, {Maraschi}, {Marcote}, {Mariotti},
  {Mart{\'{\i}}nez}, {Mazin}, {Menzel}, {Miranda}, {Mirzoyan}, {Moralejo},
  {Munar-Adrover}, {Nakajima}, {Niedzwiecki}, {Nilsson}, {Nishijima}, {Noda},
  {Orito}, {Overkemping}, {Paiano}, {Palatiello}, {Paneque}, {Paoletti},
  {Paredes}, {Paredes-Fortuny}, {Persic}, {Poutanen}, {Moroni}, {Prandini},
  {Puljak}, {Reinthal}, {Rhode}, {Rib{\'o}}, {Rico}, {Garcia}, {R{\"u}gamer},
  {Saito}, {Saito}, {Satalecka}, {Scalzotto}, {Scapin}, {Schultz}, {Schweizer},
  {Shore}, {Sillanp{\"a}{\"a}}, {Sitarek}, {Snidaric}, {Sobczynska}, {Spanier},
  {Stamatescu}, {Stamerra}, {Steinbring}, {Storz}, {Strzys}, {Takalo},
  {Takami}, {Tavecchio}, {Temnikov}, {Terzi{\'c}}, {Tescaro}, {Teshima},
  {Thaele}, {Tibolla}, {Torres}, {Toyama}, {Treves}, {Uellenbeck}, {Vogler},
  {Zanin}, {Kadler}, {Schulz}, {Ros}, {Bach}, {Krau{\ss}}, \&
  {Wilms}}]{ReferenceVariability4}
{Aleksi{\'c}}, J., {Ansoldi}, S., {Antonelli}, L.~A., {et~al.} 2014, Science,
  346, 1080

\bibitem[{{Barkov} {et~al.}(2010){Barkov}, {Aharonian}, \&
  {Bosch-Ramon}}]{2010ApJ...724.1517B}
{Barkov}, M.~V., {Aharonian}, F.~A., \& {Bosch-Ramon}, V. 2010, \apj, 724, 1517

\bibitem[{{Becerra Gonz{\'a}lez} {et~al.}(2020){Becerra Gonz{\'a}lez},
  {Acosta-Pulido}, {Boschin}, {Clavero}, {Otero-Santos}, {Carballo-Bello}, \&
  {Dom{\'\i}nguez-Palmero}}]{PepaInPrep}
{Becerra Gonz{\'a}lez}, J., {Acosta-Pulido}, J.~A., {Boschin}, W., {et~al.}
  2020, arXiv e-prints, arXiv:2010.14532

\bibitem[{{Bednarek} \& {Protheroe}(1997)}]{1997MNRAS.287L...9B}
{Bednarek}, W. \& {Protheroe}, R.~J. 1997, \mnras, 287, L9

\bibitem[{{Beskin} {et~al.}(1992){Beskin}, {Istomin}, \& {Parev}}]{Beskin}
{Beskin}, V.~S., {Istomin}, Y.~N., \& {Parev}, V.~I. 1992, \sovast, 36, 642

\bibitem[{{Blandford} \& {Levinson}(1995)}]{BlandfordLevinson}
{Blandford}, R.~D. \& {Levinson}, A. 1995, \apj, 441, 79

\bibitem[{{Blandford} \& {Znajek}(1977)}]{BlandfordZnajek}
{Blandford}, R.~D. \& {Znajek}, R.~L. 1977, \mnras, 179, 433

\bibitem[{{Broderick} \& {Tchekhovskoy}(2015)}]{Broderick}
{Broderick}, A.~E. \& {Tchekhovskoy}, A. 2015, \apj, 809, 97

\bibitem[{{Celotti} {et~al.}(1997){Celotti}, {Padovani}, \&
  {Ghisellini}}]{1997MNRAS.286..415C}
{Celotti}, A., {Padovani}, P., \& {Ghisellini}, G. 1997, \mnras, 286, 415

\bibitem[{{Chen} \& {Yuan}(2020)}]{Chen2}
{Chen}, A.~Y. \& {Yuan}, Y. 2020, \apj, 895, 121

\bibitem[{{Chen} {et~al.}(2018){Chen}, {Yuan}, \& {Yang}}]{Chen}
{Chen}, A.~Y., {Yuan}, Y., \& {Yang}, H. 2018, \apj, 863, L31

\bibitem[{{Cheng} {et~al.}(1986){Cheng}, {Ho}, \& {Ruderman}}]{CHR2}
{Cheng}, K.~S., {Ho}, C., \& {Ruderman}, M. 1986, \apj, 300, 522

\bibitem[{{Crinquand} {et~al.}(2020){Crinquand}, {Cerutti}, {Philippov},
  {Parfrey}, \& {Dubus}}]{Crinquand}
{Crinquand}, B., {Cerutti}, B., {Philippov}, A.~e., {Parfrey}, K., \& {Dubus},
  G. 2020, \prl, 124, 145101

\bibitem[{{Dermer} \& {Giebels}(2016)}]{Dermer}
{Dermer}, C.~D. \& {Giebels}, B. 2016, Comptes Rendus Physique, 17, 594

\bibitem[{{Ford} {et~al.}(2018){Ford}, {Keenan}, \& {Medvedev}}]{Ford}
{Ford}, A.~L., {Keenan}, B.~D., \& {Medvedev}, M.~V. 2018, \prd, 98, 063016

\bibitem[{{Foschini}(2017)}]{BlazarReview}
{Foschini}, L. 2017, Frontiers in Astronomy and Space Sciences, 4, 6

\bibitem[{{Foschini} {et~al.}(2013){Foschini}, {Bonnoli}, {Ghisellini},
  {Tagliaferri}, {Tavecchio}, \& {Stamerra}}]{ReferenceVariability3}
{Foschini}, L., {Bonnoli}, G., {Ghisellini}, G., {et~al.} 2013, \aap, 555, A138

\bibitem[{{Ghisellini} {et~al.}(2017){Ghisellini}, {Righi}, {Costamante}, \&
  {Tavecchio}}]{FermiBlazarSequence}
{Ghisellini}, G., {Righi}, C., {Costamante}, L., \& {Tavecchio}, F. 2017,
  \mnras, 469, 255

\bibitem[{{Gould}(1979)}]{Gould}
{Gould}, R.~J. 1979, \aap, 76, 306

\bibitem[{{Hirotani}(2005)}]{Hirotani}
{Hirotani}, K. 2005, Advances in Space Research, 35, 1085

\bibitem[{{Hirotani}(2018)}]{2018Galax...6..122H}
{Hirotani}, K. 2018, Galaxies, 6, 122

\bibitem[{{Hirotani} \& {Okamoto}(1998)}]{1998Hirotani}
{Hirotani}, K. \& {Okamoto}, I. 1998, \apj, 497, 563

\bibitem[{{Hirotani} \& {Pu}(2016)}]{HirotaniPu1}
{Hirotani}, K. \& {Pu}, H.-Y. 2016, \apj, 818, 50

\bibitem[{{Hirotani} {et~al.}(2016){Hirotani}, {Pu}, {Lin}, {Chang}, {Inoue},
  {Kong}, {Matsushita}, \& {Tam}}]{HirotaniPu2}
{Hirotani}, K., {Pu}, H.-Y., {Lin}, L. C.-C., {et~al.} 2016, \apj, 833, 142

\bibitem[{{Hirotani} {et~al.}(2017){Hirotani}, {Pu}, {Lin}, {Kong},
  {Matsushita}, {Asada}, {Chang}, \& {Tam}}]{2017Hiro}
{Hirotani}, K., {Pu}, H.-Y., {Lin}, L. C.-C., {et~al.} 2017, \apj, 845, 77

\bibitem[{{Hunter}(2007)}]{Hunter}
{Hunter}, J.~D. 2007, Computing in Science and Engineering, 9, 90

\bibitem[{{Jones}(1968)}]{Jones}
{Jones}, F.~C. 1968, Physical Review, 167, 1159

\bibitem[{{Katsoulakos} \& {Rieger}(2020)}]{2020ApJ...895...99K}
{Katsoulakos}, G. \& {Rieger}, F.~M. 2020, \apj, 895, 99

\bibitem[{{Katsoulakos} {et~al.}(2020){Katsoulakos}, {Rieger}, \&
  {Reville}}]{2020ApJ...899L...7K}
{Katsoulakos}, G., {Rieger}, F.~M., \& {Reville}, B. 2020, \apjl, 899, L7

\bibitem[{{Kisaka} {et~al.}(2020){Kisaka}, {Levinson}, \&
  {Toma}}]{2020arXiv200702838K}
{Kisaka}, S., {Levinson}, A., \& {Toma}, K. 2020, \apj, 902, 80

\bibitem[{{Levinson}(2000)}]{2000PhRvL..85..912L}
{Levinson}, A. 2000, \prl, 85, 912

\bibitem[{{Levinson} \& {Cerutti}(2018)}]{LevinsonCerutti}
{Levinson}, A. \& {Cerutti}, B. 2018, \aap, 616, A184

\bibitem[{{Levinson} \& {Rieger}(2011)}]{LR}
{Levinson}, A. \& {Rieger}, F. 2011, \apj, 730, 123

\bibitem[{Levinson \& Segev(2017)}]{Levinson}
Levinson, A. \& Segev, N. 2017, Phys. Rev. D, 96, 123006

\bibitem[{{Liao}(2018)}]{ReferenceVariability7}
{Liao}, N.-h. 2018, Galaxies, 6, 68

\bibitem[{{Liodakis}(2018)}]{Liodakis}
{Liodakis}, I. 2018, \aap, 616, A93

\bibitem[{{Lovelace} {et~al.}(1979){Lovelace}, {MacAuslan}, \&
  {Burns}}]{Lovelace}
{Lovelace}, R.~V.~E., {MacAuslan}, J., \& {Burns}, M. 1979, in American
  Institute of Physics Conference Series, Vol.~56, Particle Acceleration
  Mechanisms in Astrophysics, ed. J.~{Arons}, C.~{McKee}, \& C.~{Max}, 399--415

\bibitem[{{MAGIC Collaboration} {et~al.}(2020){MAGIC Collaboration}, {Acciari},
  {Ansoldi}, {Antonelli}, {Babi{\'c}}, {Banerjee}, {Barres de Almeida},
  {Barrio}, {Becerra Gonz{\'a}lez}, {Bednarek}, {Bernardini}, {Berti},
  {Besenrieder}, {Bhattacharyya}, {Bigongiari}, {Blanch}, {Bonnoli}, {Busetto},
  {Carosi}, {Ceribella}, {Cikota}, {Colak}, {Colin}, {Colombo}, {Contreras},
  {Cortina}, {Covino}, {D'Elia}, {da Vela}, {Dazzi}, {de Angelis}, {de Lotto},
  {Delfino}, {Delgado}, {di Pierro}, {Do Souto Espi{\~n}era}, {Dom{\'\i}nguez},
  {Dominis Prester}, {Doro}, {Fallah Ramazani}, {Fattorini},
  {Fern{\'a}ndez-Barral}, {Ferrara}, {Fidalgo}, {Foffano}, {Fonseca}, {Font},
  {Fruck}, {Galindo}, {Gallozzi}, {Garc{\'\i}a L{\'o}pez}, {Garczarczyk},
  {Gasparyan}, {Gaug}, {Giammaria}, {Godinovi{\'c}}, {Guberman}, {Hadasch},
  {Hahn}, {Hassan}, {Herrera}, {Hoang}, {Hrupec}, {Inoue}, {Ishio}, {Iwamura},
  {Kubo}, {Kushida}, {Kuve{\v{z}}di{\'c}}, {Lamastra}, {Lelas}, {Leone},
  {Lindfors}, {Lombardi}, {Longo}, {L{\'o}pez}, {L{\'o}pez-Oramas}, {Machado de
  Oliveira Fraga}, {Maggio}, {Majumdar}, {Makariev}, {Mallamaci}, {Maneva},
  {Manganaro}, {Maraschi}, {Mariotti}, {Mart{\'\i}nez}, {Masuda}, {Mazin},
  {Minev}, {Miranda}, {Mirzoyan}, {Molina}, {Moralejo}, {Moreno}, {Moretti},
  {Munar-Adrover}, {Neustroev}, {Niedzwiecki}, {Nievas Rosillo}, {Nigro},
  {Nilsson}, {Ninci}, {Nishijima}, {Noda}, {Nogu{\'e}s}, {Paiano}, {Palacio},
  {Paneque}, {Paoletti}, {Paredes}, {Pedaletti}, {Pe{\~n}il}, {Peresano},
  {Persic}, {Prada Moroni}, {Prandini}, {Puljak}, {Garcia}, {Rib{\'o}}, {Rico},
  {Righi}, {Rugliancich}, {Saha}, {Sahakyan}, {Saito}, {Satalecka},
  {Schweizer}, {Sitarek}, {{\v{S}}nidari{\'c}}, {Sobczynska}, {Somero},
  {Stamerra}, {Strzys}, {Suri{\'c}}, {Tavecchio}, {Temnikov}, {Terzi{\'c}},
  {Teshima}, {Torres-Alb{\`a}}, {Tsujimoto}, {van Scherpenberg}, {Vanzo},
  {Vazquez Acosta}, {Vovk}, {Will}, {Zari{\'c}}, {Fact Collaboration},
  {Arbet-Engels}, {Baack}, {Balbo}, {Biland}, {Blank}, {Bretz}, {Bruegge},
  {Bulinski}, {Buss}, {Doerr}, {Dorner}, {Einecke}, {Elsaesser}, {Hildebrand},
  {Linhoff}, {Mannheim}, {Mueller}, {Neise}, {Neronov}, {Noethe}, {Paravac},
  {Rhode}, {Schleicher}, {Schulz}, {Sedlaczek}, {Shukla}, {Sliusar}, {von
  Willert}, {Walter}, {Wendel}, {Tramacere}, {Lien}, {Perri}, {Verrecchia},
  {Armas Padilla}, {Leto}, {L{\"a}hteenm{\"a}ki}, {Tornikoski}, \&
  {Tammi}}]{JBG_Mrk501_bump}
{MAGIC Collaboration}, {Acciari}, V.~A., {Ansoldi}, S., {et~al.} 2020, \aap,
  637, A86

\bibitem[{{MAGIC Collaboration} {et~al.}(2018){MAGIC Collaboration}, {Ansoldi},
  {Antonelli}, {Arcaro}, {Baack}, {Babi{\'c}}, {Banerjee}, {Bangale}, {Barres
  de Almeida}, {Barrio}, {Becerra Gonz{\'a}lez}, {Bednarek}, {Bernardini},
  {Berse}, {Berti}, {Bhattacharyya}, {Bigongiari}, {Biland}, {Blanch},
  {Bonnoli}, {Carosi}, {Ceribella}, {Chatterjee}, {Colak}, {Colin}, {Colombo},
  {Contreras}, {Cortina}, {Covino}, {Cumani}, {D'Elia}, {da Vela}, {Dazzi}, {de
  Angelis}, {de Lotto}, {Delfino}, {Delgado}, {di Pierro}, {Dom{\'\i}nguez},
  {Dominis Prester}, {Dorner}, {Doro}, {Einecke}, {Elsaesser}, {Fallah
  Ramazani}, {Fattorini}, {Fern{\'a}ndez-Barral}, {Ferrara}, {Fidalgo},
  {Foffano}, {Fonseca}, {Font}, {Fruck}, {Galindo}, {Gallozzi}, {Garc{\'\i}a
  L{\'o}pez}, {Garczarczyk}, {Gaug}, {Giammaria}, {Godinovi{\'c}}, {Gora},
  {Guberman}, {Hadasch}, {Hahn}, {Hassan}, {Hayashida}, {Herrera}, {Hoang},
  {Hose}, {Hrupec}, {Ishio}, {Konno}, {Kubo}, {Kushida}, {Lamastra}, {Lelas},
  {Leone}, {Lindfors}, {Lombardi}, {Longo}, {L{\'o}pez}, {Maggio}, {Majumdar},
  {Makariev}, {Maneva}, {Manganaro}, {Mannheim}, {Maraschi}, {Mariotti},
  {Mart{\'\i}nez}, {Masuda}, {Mazin}, {Mielke}, {Minev}, {Miranda}, {Mirzoyan},
  {Moralejo}, {Moreno}, {Moretti}, {Nagayoshi}, {Neustroev}, {Niedzwiecki},
  {Nievas Rosillo}, {Nigro}, {Nilsson}, {Ninci}, {Nishijima}, {Noda},
  {Nogu{\'e}s}, {Paiano}, {Palacio}, {Paneque}, {Paoletti}, {Paredes},
  {Pedaletti}, {Pe{\~n}il}, {Peresano}, {Persic}, {Pfrang}, {Prada Moroni},
  {Prandini}, {Puljak}, {Garcia}, {Reichardt}, {Rhode}, {Rib{\'o}}, {Rico},
  {Righi}, {Rugliancich}, {Saha}, {Saito}, {Satalecka}, {Schweizer}, {Sitarek},
  {{\v{S}}nidari{\'c}}, {Sobczynska}, {Stamerra}, {Strzys}, {Suri{\'c}},
  {Takahashi}, {Tavecchio}, {Temnikov}, {Terzi{\'c}}, {Teshima},
  {Torres-Alb{\`a}}, {Tsujimoto}, {Vanzo}, {Vazquez Acosta}, {Vovk}, {Ward},
  {Will}, {Zari{\'c}}, {Glawion}, {Takalo}, \&
  {Jormanainen}}]{ReferenceVariability6}
{MAGIC Collaboration}, {Ansoldi}, S., {Antonelli}, L.~A., {et~al.} 2018, \aap,
  617, A91

\bibitem[{{Mahadevan}(1997)}]{Mahadevan}
{Mahadevan}, R. 1997, \apj, 477, 585

\bibitem[{{Maraschi} {et~al.}(1992){Maraschi}, {Ghisellini}, \&
  {Celotti}}]{Maraschi}
{Maraschi}, L., {Ghisellini}, G., \& {Celotti}, A. 1992, \apjl, 397, L5

\bibitem[{{Moles} {et~al.}(1987){Moles}, {Masegosa}, \& {del Olmo}}]{Moles}
{Moles}, M., {Masegosa}, J., \& {del Olmo}, A. 1987, \aj, 94, 1143

\bibitem[{{Narayan} \& {Yi}(1995)}]{NY}
{Narayan}, R. \& {Yi}, I. 1995, \apj, 452, 710

\bibitem[{{Neronov} \& {Aharonian}(2007)}]{Neronov2}
{Neronov}, A. \& {Aharonian}, F.~A. 2007, \apj, 671, 85

\bibitem[{{Neronov} {et~al.}(2009){Neronov}, {Semikoz}, \& {Tkachev}}]{Neronov}
{Neronov}, A.~Y., {Semikoz}, D.~V., \& {Tkachev}, I.~I. 2009, New Journal of
  Physics, 11, 065015

\bibitem[{{Petropoulou} {et~al.}(2019){Petropoulou}, {Yuan}, {Chen}, \&
  {Mastichiadis}}]{Petropoulou}
{Petropoulou}, M., {Yuan}, Y., {Chen}, A.~Y., \& {Mastichiadis}, A. 2019, \apj,
  883, 66

\bibitem[{{Pian} {et~al.}(2005){Pian}, {Falomo}, \& {Treves}}]{Pian}
{Pian}, E., {Falomo}, R., \& {Treves}, A. 2005, \mnras, 361, 919

\bibitem[{{Pittori} {et~al.}(2018){Pittori}, {Lucarelli}, {Verrecchia},
  {Raiteri}, {Villata}, {Vittorini}, {Tavani}, {Puccetti}, {Perri},
  {Donnarumma}, {Vercellone}, {Acosta-Pulido}, {Bachev}, {Ben{\'\i}tez},
  {Borman}, {Carnerero}, {Carosati}, {Chen}, {Ehgamberdiev}, {Goded},
  {Grishina}, {Hiriart}, {Hsiao}, {Jorstad}, {Kimeridze}, {Kopatskaya},
  {Kurtanidze}, {Kurtanidze}, {Larionov}, {Larionova}, {Marscher},
  {Mirzaqulov}, {Morozova}, {Nilsson}, {Samal}, {Sigua}, {Spassov},
  {Strigachev}, {Takalo}, {Antonelli}, {Bulgarelli}, {Cattaneo},
  {Colafrancesco}, {Giommi}, {Longo}, {Morselli}, \&
  {Paoletti}}]{ReferenceVariability5}
{Pittori}, C., {Lucarelli}, F., {Verrecchia}, F., {et~al.} 2018, \apj, 856, 99

\bibitem[{{Ptitsyna} \& {Neronov}(2016)}]{Ptitsyna}
{Ptitsyna}, K. \& {Neronov}, A. 2016, \aap, 593, A8

\bibitem[{{Rieger}(2019)}]{RiegerVariability}
{Rieger}, F. 2019, Galaxies, 7, 28

\bibitem[{{Stocke} {et~al.}(2011){Stocke}, {Danforth}, \& {Perlman}}]{Stocke}
{Stocke}, J.~T., {Danforth}, C.~W., \& {Perlman}, E.~S. 2011, \apj, 732, 113

\bibitem[{{Svensson}(1987)}]{Svensson}
{Svensson}, R. 1987, \mnras, 227, 403

\bibitem[{{Tang} {et~al.}(2008){Tang}, {Takata}, {Jia}, \& {Cheng}}]{Tang}
{Tang}, A. P.~S., {Takata}, J., {Jia}, J.~J., \& {Cheng}, K.~S. 2008, \apj,
  676, 562

\bibitem[{{Urry} \& {Padovani}(1995)}]{Urry}
{Urry}, C.~M. \& {Padovani}, P. 1995, \pasp, 107, 803

\bibitem[{{Vincent} \& {Lebohec}(2010)}]{2010Vincent}
{Vincent}, S. \& {Lebohec}, S. 2010, \mnras, 409, 1183

\bibitem[{{Wendel} {et~al.}(2017){Wendel}, {Glawion}, {Shukla}, \&
  {Mannheim}}]{Wendel17}
{Wendel}, C., {Glawion}, D., {Shukla}, A., \& {Mannheim}, K. 2017, in American
  Institute of Physics Conference Series, Vol. 1792, 6th International
  Symposium on High Energy Gamma-Ray Astronomy, 050026

\bibitem[{{Wilms} {et~al.}(2000){Wilms}, {Allen}, \& {McCray}}]{Wilms}
{Wilms}, J., {Allen}, A., \& {McCray}, R. 2000, \apj, 542, 914

\bibitem[{{Zdziarski}(1988)}]{Zdz}
{Zdziarski}, A.~A. 1988, \apj, 335, 786

\end{thebibliography}

\begin{appendix}

\section{Derivation of the kinetic equations}\label{AppendixSectionDescriptionViaKineticEquations}

As an attachment to Subsect. \ref{SubsectionDescriptionViaKineticEquations}, here we give a precise derivation of the kinetic equations. To quantify the change rates of the spectral number densities of the photons and of the electrons, it is necessary to introduce the spectral interaction rates of IC scattering and of PP\@. We do this in line with \citet{Zdz}.

Given an IC scattering event of an incident electron with original energy $\gamma \gg 1$ off a photon with energy $x \ll 1$ that results in a down-scattered electron with final energy $\gamma'$ and a highly energetic photon with energy $x_\gamma$. Energy conservation yields $\gamma \approx \gamma + x = \gamma' + x_\gamma$. From IC scattering kinematics, it follows
\begin{equation}
\gamma x > E_{\ast}
\label{EquationICScatteringKinematics}
\end{equation}
with the abbreviation $E_{\ast} = (\gamma / \gamma' - 1)/4$ \citep[cf.][]{Zdz}. This ensures that the photon energy has a maximum possible value $x_{\gamma,\,\mathrm{max}}(\gamma,x) := (4 \gamma^2 x) / (1 + 4 \gamma x)$. This function is $< \gamma$ for all realistic values of $\gamma$, which reflects that the electron cannot transfer all its energy to the photon. Moreover, we remark that $x_{\gamma,\,\mathrm{max}}(\gamma,x)$ increases with increasing $\gamma$ as well as with increasing $x$. Hence, if $N$ has an upper cut-off at $\gamma_{\rm{i},\,0}$ and $n_0$ has an upper cut-off at $x_0$, then the distribution of highly energetic photons vanishes above $x_{\gamma,\,\mathrm{max}}(\gamma_{\rm{i},\,0},x_0)$ and is non-vanishing below this.

The spectral IC scattering interaction rate $C$ for such events on the field of low-energy photons with spectral number density $n_0$ is given by \citet{Zdz} as an approximation of the exact one found by \citet{Jones},
\begin{equation}
C(\gamma,\gamma') = \int_{\max(x_1, \, E_{\ast}/\gamma)}^{x_0} n_0(x) \frac{3 \sigma_{\rm{Th}} c}{4 \gamma^2 x} \left[ r + (2-r) \frac{E_{\ast}}{\gamma x} - 2 \left( \frac{E_{\ast}}{\gamma x} \right)^2 - 2 \frac{E_{\ast}}{\gamma x} \ln{\frac{\gamma x}{E_{\ast}}} \right] \, \mathrm{d}x
\label{EquationSpectralICScatteringRate}
\end{equation}
Here it is $r = (\gamma / \gamma' + \gamma' / \gamma)/2$ and the second argument in the lower integration border is due to Ineq. \ref{EquationICScatteringKinematics}.
$N(\gamma) C(\gamma,\gamma')$ is the number of IC scattering events with original energy $\gamma$ and final energy $\gamma'$ per unit time, per unit space volume, per unit original energy, and per unit final energy.

Substituting $\gamma'$ with $\gamma - x_\gamma$ in $C$, we realise that $C(\gamma,\gamma - x_\gamma)$ describes the probability for an IC scattering event with original electron energy $\gamma$ to produce a photon with energy $x_\gamma$, per unit time, and per unit $x_\gamma$-interval. $N(\gamma) C(\gamma,\gamma - x_\gamma)$ stands for the number of IC scattering events with original electron energy $\gamma$ that result in a photon with energy $x_\gamma$, per unit time, per unit space volume, per unit $\gamma$-interval, and per unit $x_\gamma$-interval.

Now, we consider a collision of a highly energetic photon of energy $x_\gamma$ with a photon of energy $x$, resulting in the PP of an electron of energy $\gamma$ and a positron with energy $x_\gamma + x - \gamma \approx x_\gamma - \gamma$. Based on \citet{Agaronyan}, \citet{Zdz} gave the corresponding kinematic relationship 
\begin{equation}
x_{\gamma} x > E_{\ast} > 1
\label{EquationPPKinematics}
\end{equation}
where $E_{\ast}=x_{\gamma}^2/(4 \gamma (x_{\gamma}-\gamma))$ as well as the spectral PP interaction rate $P$, again for a photon field $n_0$,
\begin{equation}
P(x_{\gamma},\gamma) = \int_{\max(x_1, \, E_{\ast}/\gamma)}^{x_0} n_0(x) \frac{3 \sigma_{\rm{Th}} c}{4 x_{\gamma}^2 x} \left[ r - (2+r) \frac{E_{\ast}}{x_{\gamma} x} + 2 \left( \frac{E_{\ast}}{x_{\gamma} x} \right)^2 + 2 \frac{E_{\ast}}{x_{\gamma} x} \ln{\frac{x_{\gamma} x}{E_{\ast}}} \right] \, \mathrm{d}x
\label{EquationSpectralPPRate}
\end{equation}
Here it is $r = (\gamma / (x_{\gamma}-\gamma) + (x_{\gamma}-\gamma) / \gamma)/2$ and the second argument in the lower integration border is due to the first part of Ineq. \ref{EquationPPKinematics}.
$n_\gamma(x_\gamma) P(x_\gamma,\gamma)$ is the number of PP events with photon energy $x_\gamma$ that create an electron with energy $\gamma$ per unit time, per unit space volume, per unit photon energy, and per unit electron energy.

To infer the kinetic equation of the electrons, we have to specify all relevant sinks and sources of the electron energy distribution in terms of change rates of the electron spectral number density $N$:
\begin{itemize}
	\item The spectral injection rate $\dot N_{\rm i}(\gamma)$ of electrons directly enters the kinetic equation.
	\item Electrons leave the interaction region on the escape timescale $T_{\rm{e\,esc}}$. This is expressed with the term ${N(\gamma)}/{T_{\rm{e\,esc}}(\gamma)}$.
	\item An electron of original energy $\gamma$ loses energy by becoming IC down-scattered to a lower energy $\gamma' < \gamma$, resulting in
	\begin{itemize}
		\item a sink at $\gamma$. We remark, however, that for given $x$ and $\gamma$, the minimum possible value of $\gamma'$ is not equal to 1 but has a lower limit at $\gamma'_{\rm{IC,\,min}}(\gamma,x) := \gamma/(1 + 4 x \gamma)$ as a result of scattering kinematics. From Ineq. \ref{EquationICScatteringKinematics} the limitation $\gamma' > \gamma'_{\rm{IC,\,min}}$ follows. It always is $\gamma'_{\rm{IC,\,min}} > 1$, reflecting the fact that the electron retains an amount of kinetic energy in every case. $\gamma'_{\rm{IC,\,min}}$ decreases with increasing $x$. Now, if the low-energy photons are not mono-energetic, but are distributed in energy space with an upper cut-off of $n_0(x)$ at some $x_0$, then the value of $\gamma'$ that can at least be reached for an electron with original energy $\gamma$, is $\gamma'_{\rm{IC,\,min}}(\gamma,x_0)$. Therefore we describe the number of IC scattering events with original energy $\gamma$ and any allowed final energy $\gamma'$ per unit time, per unit space volume, and per unit original energy by $\int ^{\gamma}_{\gamma'_{\rm{IC,\,min}}(\gamma,x_0)} N(\gamma) C(\gamma,\gamma') \, \rm{d}\gamma'$ \footnote{The lower integration border is in contrast to the one mistakenly used by \citet{Zdz} in the second term on the right-hand side of his Eq.~1 as well as in the first integral in A21.}.
		\item a source at $\gamma'$. Though, we can infer from Ineq. \ref{EquationICScatteringKinematics} that a final electron energy $\gamma'$ cannot be reached from any original electron energy $\gamma$. For $\gamma' \geq 1/(4 x)$ (scattering into the KN regime), the original electron energy is indeed unbounded. However, for $\gamma' < 1/(4 x)$ (scattering into the Thomson regime), the original electron energy has to be lower than\\$\gamma'/(1 - 4 x \gamma')$
		, which increases with increasing $x$. Again, if the low-energy photon spectral number density is extended with an upper cut-off at $x_0$, then the upper boundary of $\gamma$ is expressed by $\gamma'/(1 - 4 x_0 \gamma')$. In what follows we define
        \begin{equation}
        \gamma_{\mathrm{IC,\,max}}(\gamma',x) := \left\{
        \begin{array}{ll}
        \gamma'/(1 - 4 x \gamma')	& \mathrm{for} \; \; \gamma' < 1/(4 x) \mathrm{,} \\
        + \infty									& \mathrm{for} \; \; \gamma' \geq 1/(4 x)\mathrm{.}
        \end{array}
        \right.
        \label{EquationgammaICmax}
        \end{equation}
        With this, we can quantify the number of IC scattering events with final energy $\gamma'$ and any permitted original energy $\gamma$ per unit time, per unit space volume, and per unit final energy by $\int _{\gamma'}^{\gamma_{\rm{IC,\,max}}(\gamma',x_0)} N(\gamma) C(\gamma,\gamma') \, \rm{d}\gamma$ \footnote{The upper integration border is in contrast to the one erroneously used by \citet{Zdz} in the third term on the right-hand side of his Eq.~1.}. When this term is plugged into the kinetic equation, we have to exchange $\gamma$ with $\gamma'$ in this term because the kinetic equation refers to the variable $\gamma$. 
	\end{itemize}
	\item PP of an electron with energy $\gamma$ is a source at this energy. However, PP on a photon with energy $x$ that produces an electron with energy $\gamma$ is possible only if the photon energy $x_\gamma$ exceeds the PP threshold $x_{\gamma,\,\rm{PP,\,th}}(\gamma,x) := \gamma/(1 - (1/(4 \gamma x)))$. This threshold can be inferred from Ineq. \ref{EquationPPKinematics}. $x_{\gamma,\,\rm{PP,\,th}}$ adopts finite and positive values only for $\gamma > 1/(4 x)$, expressing that electrons cannot be pair-produced below $1/(4 x)$. In this range, $x_{\gamma,\,\rm{PP,\,th}}$ has a minimum at $\gamma = 1/(2 x)$. This minimum corresponds to $x_{\gamma,\,\rm{PP,\,th}}(1/(2 x),x) = 1/x$, which is the well-known set-in of PP\@. For $\gamma > 1/(4 x)$, it is of course $x_{\gamma,\,\rm{PP,\,th}} > \gamma$. Furthermore, we can show that $x_{\gamma,\,\rm{PP,\,th}}$ decreases with increasing $x$. Hence, if the low-energy photons are not mono-energetic, but are distributed with $n_0(x)$, which vanishes above an upper cut-off $x_0$, then PP is possible as soon as $x_\gamma > x_{\gamma,\,\rm{PP,\,th}}(\gamma,x_0)$. The number of kinematically allowed PP events that create an electron with energy $\gamma$ per unit time, per unit space volume, and per unit electron energy, is accordingly expressed by $\int _{x_{\gamma,\,\rm{PP,\,th}}(\gamma,x_0)}^{\infty} n_\gamma(x_\gamma) P(x_\gamma,\gamma) \, {\rm{d}}x_\gamma$. As $P$ is normalised to the production of only one electron, this term has to be multiplied by 2 in the kinetic equation.
\end{itemize}
The electron kinetic equation is obtained by setting the rate of change $\dot N(\gamma)$ of the electron spectral number density equal to the sum of all source terms subtracted by all sinks,
\begin{equation}
\dot N(\gamma) = \dot N_{\mathrm{i}}(\gamma) - N(\gamma) \left( \frac{1}{T_{\rm{e\,esc}}(\gamma)} + \int ^{\gamma}_{\gamma'_{\rm{IC,\,min}}(\gamma,x_0)} C(\gamma,\gamma') \, \mathrm{d}\gamma' \right) + \int _{\gamma}^{\gamma'_{\mathrm{IC,\,max}}(\gamma,x_0)} N(\gamma') C(\gamma',\gamma) \, \mathrm{d}\gamma' + 2 \cdot \int _{x_{\gamma,\,\mathrm{PP,\,th}}(\gamma,x_0)}^{\infty} n_\gamma(x_\gamma) P(x_\gamma,\gamma) \, \mathrm{d}x_\gamma
\label{EquationRateOfChangeElectrons}
\end{equation}
In what follows the first term of this equation is called electron spectral injection rate, the part within round brackets is called electron spectral loss rate (which is the sum of the spectral escape rate and of the spectral IC down-scattering rate to lower energies), and the sum of the third and fourth term are called electron spectral production rate (which is the sum of the spectral IC down-scattering rate from higher energies and of the spectral PP rate)\footnote{We note that the dimension of the electron spectral injection rate and of the electron spectral production rate is $[\gamma]^{-1} [t]^{-1} [l]^{-3}$, respectively, whereas that of the electron spectral loss rate is $[\gamma]^{-1} [t]^{-1}$.}.

Analogously, we identify all source and sink terms that determine the rate of change $\dot n_\gamma(x_\gamma)$ of the highly energetic photon spectral number density:
\begin{itemize}
	\item The injection term $\dot n_{\gamma,\,\rm i}(x_\gamma)$ has to be included.
	\item The spectral loss rate due to escape from the interaction volume is described by ${n_\gamma(x_\gamma)}/{T_{\rm{ph\,esc}}(x_\gamma)}$.
	\item PP is a sink of the highly energetic photon distribution at the energy $x_\gamma$ of the incident photons. For given $x$ and $x_\gamma$, the energies $\gamma$ that can be reached by the created electrons, are limited due to the interaction kinematics Ineq. \ref{EquationPPKinematics}. The electron energy $\gamma$ that can be obtained obeys $\gamma_{\rm{PP,\,min}} < \gamma < \gamma_{\rm{PP,\,max}}$ with the limits $\gamma_{\rm{PP,\,min}}(x_\gamma,x) := x_\gamma(1 - (1 - 1 /(x_\gamma x))^{1/2})/2$ and $\gamma_{\rm{PP,\,max}}(x_\gamma,x) :=\\ x_\gamma(1 + (1 - 1 /(x_\gamma x))^{1/2})/2$. With increasing $x$, the limit $\gamma_{\rm{PP,\,min}}$ decreases and $\gamma_{\rm{PP,\,max}}$ increases. Consequently, if $n_0$ is extended and non-vanishing only below and at an upper cut-off $x_0$, then the electron energies that are kinematically allowed satisfy $\gamma_{\rm{PP,\,min}}(x_\gamma,x_0) < \gamma < \gamma_{\rm{PP,\,max}}(x_\gamma,x_0)$. With this in mind, the number of PP events with incident photon energy $x_\gamma$ and any possible electron energy per unit time, per unit space volume, and per unit $x_\gamma$-interval can be written as $\int ^{\gamma_{\rm{PP,\,max}}(x_\gamma,x_0)}_{\gamma_{\rm{PP,\,min}}(x_\gamma,x_0)} n_\gamma(x_\gamma) P(x_\gamma,\gamma) \, \rm{d}\gamma$.
	\item The IC up-scattering of photons of energy $x$ to energy $x_\gamma$ is a source of photons. The energy $x_\gamma$ can, however, only be reached, when the original electron energy $\gamma$ exceeds a threshold $\gamma_{\rm{IC,\,th}}$. This threshold is defined by $\gamma_{\rm{IC,\,th}}(x_\gamma,x) := x_\gamma(1 + (1 + 1 /(x_\gamma x))^{1/2})/2$, as can be inferred from Ineq. \ref{EquationICScatteringKinematics}, and is $> x_\gamma$ for all realistic $x_\gamma$. The constraint by this threshold is again the result of the electron not being able to transfer all its energy to the photon, and hence the original electron energy has to exceed $x_\gamma$ by $\gamma'$, which cannot fall below the value $\gamma_{\rm{IC,\,th}} - x_\gamma$. We note that $\gamma_{\rm{IC,\,th}}$ decreases with increasing $x$. Thus, if $n_0$ is extended over a range of energies and has an upper cut-off at $x_0$, then IC scattering to the energy $x_\gamma$ sets in as soon as $\gamma > \gamma_{\rm{IC,\,th}}(x_\gamma,x_0)$. Then, $\int _{\gamma_{\rm{IC,\,th}}(x_\gamma,x_0)}^{\infty} N(\gamma) C(\gamma,\gamma-x_\gamma) \, \rm{d}\gamma$ quantifies the number of IC scattering events with any kinematically allowed original electron energy that result in a photon with energy $x_\gamma$ per unit time, per unit space volume, and per unit energy.
\end{itemize}
The highly energetic photon kinetic equation is obtained by setting the rate of change $\dot n_\gamma(x_\gamma)$ of the photon spectral number density equal to the sum of all source terms subtracted by all sinks,
\begin{equation}
\dot n_\gamma(x_\gamma) = \dot n_{\gamma,\,\mathrm{i}}(x_\gamma) - n_\gamma(x_\gamma) \left( \frac{1}{T_{\rm{ph\,esc}}(x_\gamma)} + \int ^{\gamma_{\mathrm{PP,\,max}}(x_\gamma,x_0)}_{\gamma_{\mathrm{PP,\,min}}(x_\gamma,x_0)} P(x_\gamma,\gamma) \, \mathrm{d}\gamma \right) + \underbrace{\int _{\gamma_{\mathrm{IC,\,th}}(x_\gamma,x_0)}^{\infty} N(\gamma) C(\gamma,\gamma-x_\gamma) \, \mathrm{d}\gamma}_{:= \, \dot n_{\gamma, \, \rm{IC}}(x_\gamma)}
\label{EquationRateOfChangeHEPs}
\end{equation}
In what follows the first term of this equation is called photon spectral injection rate, the part within round brackets is called photon spectral loss rate (which is the sum of the spectral escape rate and of the spectral pair absorption rate), and the last term $\dot n_{\gamma, \, \rm{IC}}$ is called photon spectral production rate\footnote{We remark that the dimension of the photon spectral injection rate and of the photon spectral production rate is $[x_{\gamma}]^{-1} [t]^{-1} [l]^{-3}$, respectively, whereas that of the photon spectral loss rate is $[x_{\gamma}]^{-1} [t]^{-1}$.}.

We assume the spectral number densities to be in steady state. Thus, the respective rates of change vanish in Eq.~\ref{EquationRateOfChangeElectrons} and \ref{EquationRateOfChangeHEPs}. Then, we can solve the respective equation for the respective spectral number density in the second term on the right-hand side of the respective equation. This yields
\begin{equation}
N(\gamma) = \frac{\dot N_{\mathrm{i}}(\gamma) + \int _{\gamma}^{\gamma'_{\mathrm{IC,\,max}}(\gamma,x_0)} N(\gamma') C(\gamma',\gamma) \, \mathrm{d}\gamma' + 2 \cdot \int _{x_{\gamma,\,\mathrm{PP,\,th}}(\gamma,x_0)}^{\infty} n_\gamma(x_\gamma) P(x_\gamma,\gamma) \, \mathrm{d}x_\gamma}{\frac{1}{T_{\rm{e\,esc}}(\gamma)} + \int ^{\gamma}_{\gamma'_{\rm{IC,\,min}}(\gamma,x_0)} C(\gamma,\gamma') \, \mathrm{d}\gamma'}
\label{EquationSpectralNumberDensityElectrons}
\end{equation}
and Eq.~\ref{EquationSpectralNumberDensityHEPs}.

\section{Discussion of numerical solution procedure}\label{AppendixSectionNumericalSolutionProcedure}

As an extension to Sect. \ref{SectionNumericalSolutionProcedure}, in this Appendix we concisely explain how explicit converging solutions for $N$ are found from iteratively solving $N(\gamma) = \mathcal{F}(N, \gamma)$ along the sequence of points $\left( \gamma_k \right)_{k = 1,\dots,\kappa}$ after prescribing the initialisation $N_{\rm{init}}(\gamma)$.

One could iterate the following way: One could determine $\mathcal{F}$ at all the points $\gamma_k$ based on the sequence $\left( N_k \right)_{j_{\rm{init}}}$. Then, one would assign the attained values to a new sequence $\left( N_k \right)_0$. In short, the initialisation step of the iteration would be described by $N_{k,\,0} = \mathcal{F}(\left( N_k \right)_{j_{\rm{init}}}, \gamma_k)$. The same technique would be pursued in the subsequent iteration steps. In each iteration step, one would compute a complete new sequence $\left( N_k \right)_j$ (with integer $j$ running along $j_{\rm{init}},\,0,\,1,\dots,\,j_{\rm{final}}$ and denoting the number of the iteration step, or equivalently, the number of the sequence of values $N_{k,\,j}$). In other words, in the $j$-th iteration step, one would determine the $\kappa$ values $N_{k,\,j} = \mathcal{F}(\left( N_k \right)_{j-1}, \gamma_k)$ based on the sequence $\left( N_k \right)_{j-1}$ of the previous iteration. Iterating would have to proceed until all points $N_{k,\,j}$ converge with a value called $N_{k,\,j_{\rm{final}}}$ (and hence the function $N(\gamma)$ converges), which is always the case according to the Banach fixed-point theorem.

\citet{Wendel17} used this iteration scheme. However, this scheme is not favoured because it is computationally inefficient. This can be seen as follows. By inspection of Eq.~\ref{EquationKineticEquation} we realise that a computation of $\mathcal{F}$ at a given point $\gamma_{k'}$ evaluates (and thus needs to know) $N$ only at points $\gamma_k \geq \gamma_{k'}$.\footnote{This observation is obvious in the case of the second term in the numerator of Eq.~\ref{EquationKineticEquation} because the integration range stretches from $\gamma$ upwards. In the case of the third term, we have to consider the values of the lower integration borders of the nested integrals. First, we make use of the fact that the lower border $x_{\gamma,\,\rm{PP,\,th}}(\gamma,x_0)$ of the outer integral is the PP threshold and thus $> \gamma$ (for all realistic values of $\gamma$). Second, as mentioned in Sect. \ref{SubsectionDescriptionViaKineticEquations}, the lower border $\gamma_{\mathrm{IC,\,th}}(x_\gamma,x_0)$ of the inner integral is always $> x_\gamma$. As an effect, the lowest value, at which $N$ has to be evaluated during the computation of the third term, is $\gamma_{\mathrm{IC,\,th}}(x_{\gamma,\,\rm{PP,\,th}}(\gamma,x_0),x_0)$, which is always $> \gamma$.} We consider the computation of a certain value $N_{k',\,j}$, in other words, the $j$-th computation of the value of $N$ at $\gamma_{k'}$. This computation is based on all the values $N_{k,\,j-1}$ with $k \geq k'$, in other words, on the sequence $\left( N_k \right)_{k = k'+1,\dots,\kappa;\,j-1}$. As long as the values with $k>k'$ have not yet converged, the computation of $N_{k',\,j}$ is based on inaccurate prerequisites and thus becomes inaccurate by itself. It is therefore reasonable to implement the following iteration scheme:

In contrast to above, where the complete sequence $\left( N_k \right)_j$ of $\kappa$ values was determined in one iteration step, we iterate point-wise. In the first step, only the value $N_{\kappa,\,j}$ at the largest $\gamma$ is iterated. Briefly written, we calculate $N_{\kappa,\,0} = \mathcal{F}(\left( N_k \right)_{j_{\rm{init}}}, \gamma_\kappa)$ and $N_{\kappa,\,j} = \mathcal{F}(\left( N_k \right)_{j-1}, \gamma_\kappa)$. Not until convergence of $N_{\kappa,\,j}$ was achieved with the final value being $N_{\kappa,\,j_{\rm{final}}}$, did we iterate at the next lower point $\gamma_{\kappa-1}$. In short, we compute $N_{\kappa-1,\,0} = \mathcal{F}(\left( N_k \right)_{j_{\rm{init}}}, \gamma_{\kappa-1})$ and $N_{\kappa-1,\,j} = \mathcal{F}(\left( N_k \right)_{j-1}, \gamma_{\kappa-1})$ until $N_{\kappa-1,\,j}$ converges with the final value being $N_{\kappa-1,\,j_{\rm{final}}}$. Then, we switch to $\gamma_{\kappa-2}$ and iterate again until convergence. This is done successively at all points until convergence was reached at the point $k=1$. This procedure is advantageous because during the iteration at a point $\gamma_{k'}$, the final converged sequence $\left( N_k \right)_{k = k'+1,\dots,\kappa;\,j_{\rm{final}}}$ of values is already known and can be used, which reduces inaccuracies at $\gamma_{k'}$.

In the range $\gamma > 1/(4 x_0)$ pairs can be produced. This is reflected in the third term in the numerator of Eq.~\ref{EquationKineticEquation} being non-zero. This makes the computation of $\mathcal{F}$ expensive in this regime. At $\gamma \leq 1/(4 x_0)$, PP does not happen and the third term vanishes, which speeds up the computations. On the other side, more iteration steps are necessary for lower values of $\gamma$ due to the decreasing energy transfer in one IC scattering event. In the deep Thomson regime (i.e.\ for $\gamma \ll 1/(4 x_0)$) Eq.~\ref{EquationRateOfChangeElectrons} in the steady state can be expressed as a continuity equation and could be solved by a differential equation solver (e.g.\ a Runge-Kutta method, cf.\ \citet{Zdz}). However, as we are mainly interested in the course of $N$ in the KN regime, we use the iteration procedure over the complete range.

\end{appendix}

\end{document}